\documentclass[a4paper,fleqn]{cas-sc}

\usepackage[numbers]{natbib}
\usepackage{bm}
\usepackage{graphicx}
\usepackage{subcaption}
\usepackage{booktabs}

\def\tsc#1{\csdef{#1}{\textsc{\lowercase{#1}}\xspace}}
\tsc{WGM}
\tsc{QE}
\tsc{EP}
\tsc{PMS}
\tsc{BEC}
\tsc{DE}

\begin{document}
\let\WriteBookmarks\relax
\def\floatpagepagefraction{1}
\def\textpagefraction{.001}
\shorttitle{A Compression-Directional Entropic Stress Method for Shock-Regularized Compressible Flow}
\shortauthors{Bonan Xu et al.}

\title [mode = title]{A Compression-Directional Entropic Stress Method for Shock-Regularized Compressible Flow}                      
\tnotemark[1,2]

	\author[1]{Bonan Xu}


    \affiliation[1]{organization={Department of Aeronautical and Aviation Engineering, The Hong Kong Polytechnic University},
	addressline={11 Yuk Choi Road}, 
	city={Hong Kong SAR},
	country={China}}

	\author[1]{Chihyung Wen}
    \cormark[1]
    \ead{chihyung.wen@polyu.edu.hk}
    
    \author[1]{Peixu Guo}

\begin{abstract}
We introduce the Compression-Directional Entropic Stress (CoDeS) method inspired by information geometric regularization. CoDeS replaces scalar multidimensional entropic pressure with a tensor stress aligned with the principal directions of compression. The stress has the form $\boldsymbol{\Pi}_{\Sigma}=\sigma\boldsymbol{M}$, where $\sigma$ is obtained from a modified-Helmholtz equation and $\boldsymbol{M}$ is constructed from the compressive eigenspace of the symmetric velocity-gradient tensor. The source is gated by volumetric and principal-strain compression, so the regularization vanishes in smooth expansion, rigid-body rotation, and ideal contacts, while recovering the compressive one-dimensional IGR mechanism at planar shocks. The same tensor stress is used in the conservative momentum flux and the stress-work energy flux. CoDeS is tested on one-, two-, and three-dimensional problems including smooth expansion, double rarefaction, the Sod shock tube, multidimensional Riemann flow, a viscous shock tube, a two-fluid triple point, a Mach-3 slot jet, and a supersonic Taylor--Green vortex. The results show that CoDeS remains inactive in expansive and contact regions, supplies localized stress at shocks, and concentrates regularization along compressive wave structures while remaining weak in shear- and vorticity-dominated regions. At matched resolutions, the three-dimensional Taylor--Green results are comparable to or more energetic than seventh-order WENO/TENO references. These results indicate that CoDeS provides a compression-selective shock regularization compatible with high-order finite-volume resolution of contacts, interfaces, shear layers, and vortical structures. All the code, case settings, and code for plotting figures of this paper are available at \url{https://github.com/xubonan/code\_for\_CoDeS}.
\end{abstract}




\maketitle

\section{Introduction}\label{sec:intro}

Discontinuities such as shock waves remain a central challenge in the numerical simulation of compressible flows. Steep gradients or discontinuities may develop even from smooth initial data in high-Mach-number flow governed by the Euler or Navier--Stokes equations. A numerical method must therefore suppress nonphysical Gibbs-type oscillations while preserving acoustic waves, vortical structures, contact discontinuities, material interfaces, and turbulent small scales. The classical shock-capturing framework is built on conservative finite-volume discretizations, upwind fluxes, approximate Riemann solvers, and nonlinear stabilization mechanisms \cite{Godunov1959,HartenLaxVanLeer1983,Roe1981,Einfeldt1988, ToroSpruceSpeares1994}. These ideas form the basis of many robust methods for hyperbolic conservation laws and compressible gas dynamics.

High-order shock-capturing methods improve accuracy away from discontinuities by combining high-order reconstruction with nonlinear adaptation near discontinuous features. Essentially non-oscillatory (ENO) and weighted ENO (WENO) schemes introduced a systematic way to avoid oscillatory stencils while maintaining high-order accuracy in smooth regions
\cite{HartenEtAl1987,ShuOsher1988,LiuOsherChan1994,JiangShu1996}. Subsequent developments, including mapped WENO, WENO-Z, monotonicity-preserving reconstruction, and targeted ENO (TENO) schemes, have improved critical-point accuracy, reduced dissipation, and enhanced robustness for compressible-flow simulations \cite{HenrickAslamPowers2005,BorgesEtAl2008,SureshHuynh1997, FuHuAdams2016,FuHuAdams2017,FuHuAdams2018}. These methods have been highly successful, but their nonlinear reconstruction or limiting procedures still reflect a compromise between stability at shocks and resolution of broadband smooth flow.

A complementary approach is to regularize the governing equations by adding artificial dissipative mechanisms that are localized near shocks or other underresolved discontinuities. The classical artificial-viscosity idea of von Neumann and Richtmyer remains influential \cite{vonNeumannRichtmyer1950}, and many modern variants use shock sensors, localized artificial diffusivity, hyperviscosity, or entropy-production indicators to reduce dissipation away from shocks \cite{DucrosEtAl1999,CookCabot2005,FiorinaLele2007,KawaiLele2008, KawaiShankarLele2010,GuermondPasquettiPopov2011,PerssonPeraire2006}. Such methods are particularly relevant for shock--turbulence interaction and high-speed turbulent flows, where excessive numerical dissipation can overwhelm the small-scale dynamics \cite{JohnsenEtAl2010,Pirozzoli2011}. The design of a shock regularization is therefore a question not only of shock stability, but also of selectivity, since the added mechanism should distinguish compressive shock-forming structures from smooth expansion, shear, and rotation.

Information geometric regularization (IGR) provides a recent PDE-level regularization strategy for Euler-type fluid models. IGR regularizes shock formation by modifying the geometry underlying the fluid motion, replacing shock singularities by smooth profiles with a controlled regularization length scale. Cao and Sch\"afer developed one-dimensional IGR formulations for barotropic and pressureless Euler-type equations and rigorously proved the existence of global smooth IGR solutions, together with their convergence to entropy solutions in the limit of vanishing regularization \cite{CaoSchaefer2023BarotropicIGR,CaoSchaefer2024PressurelessIGR}. For the compressible Euler equations with thermodynamics, Hamiltonian and metriplectic analyses have clarified the conservative and dissipative components of IGR, including its entropy-production and energy-transport properties \cite{BarhamEtAl2025HIGR}. Finite-volume IGR implementations have further shown that an IGR entropic-pressure equation can be incorporated into practical high-order finite-volume solvers and compares favorably with standard shock-capturing methods on canonical compressible-flow benchmarks \cite{RadhakrishnanEtAl2026FVMIGR}. More recently, Taylor et al. extend IGR by incorporating a Shannon-entropy/Kullback--Leibler thermodynamic constraint into the Hessian metric geometry \cite{TaylorSpiteriGaudreault2026TIGRE}. Their TIGRE model introduces an entropy-coupled anisotropic stress and an additional elliptic equation, and is designed to mitigate the thermodynamic cusp singularities observed in earlier thermodynamic IGR formulations.

The present work builds on this IGR direction, but addresses a multidimensional selectivity issue of scalar entropic-pressure regularization. In multidimensional flows, compression, expansion, shear, and rotation coexist, and a scalar isotropic regularization cannot distinguish the principal directions of local compression. Moreover, scalar invariants of the full velocity-gradient tensor can involve both strain and rotational contributions. As a result, a scalar multidimensional regularization may be activated in shear- or vorticity-dominated regions, even when the desired mechanism targets shock-forming compression. This is undesirable in flows where shocks interact with vortices, contacts, material interfaces, wall-bounded shear layers, or jet shear layers.

We introduce the Compression-Directional Entropic Stress (CoDeS) method. CoDeS retains the IGR idea of obtaining a smooth regularization amplitude from a modified-Helmholtz equation, but replaces the scalar multidimensional entropic pressure by a tensor stress aligned with the local compressive eigenspace of the symmetric velocity-gradient tensor. The scalar source is gated by net volumetric compression and by principal compressive strain, so the closure is inactive in pure expansion, pure rigid-body rotation, and ideal contacts with constant velocity and pressure. In one spatial dimension, the closure recovers the compressive part of the scalar IGR mechanism, so planar shocks remain regularized. In smooth compressible flows, the compression-aligned tensor stress yields a formal nonnegative entropy-production mechanism. These structural properties are used as continuous design principles. The present paper does not claim a full discrete entropy-stability theorem.

The CoDeS closure has three practical features. First, it requires only one scalar modified-Helmholtz solve per time-integration stage, since the tensor direction is computed locally from the symmetric velocity-gradient tensor. Second, the regularization enters through conservative momentum and energy stress fluxes, and is therefore compatible with standard finite-volume reconstruction and numerical fluxes. Third, the baseline continuous closure is separated from two optional finite-volume robustness safeguards. One is a shock-localized face-normal compression source that augments only the scalar stress-amplitude source, and the other is a boundary-localized flux fallback used only near severe nonperiodic shock-boundary interactions. This separation allows the baseline compression-directional closure to be tested separately from these optional finite-volume safeguards.

The numerical study assesses whether CoDeS achieves the intended wave-type and directional selectivity. One-dimensional smooth-expansion, double-rarefaction, and Sod shock-tube tests examine whether the method remains inactive in expansion and contacts while retaining shock regularization. A two-dimensional isentropic vortex probes long-time behavior in smooth vortical flows. A perturbed two-dimensional Riemann problem, a viscous shock-tube problem, multi-material triple-point calculations, a Mach-3 underexpanded slot jet, and a supersonic Taylor--Green vortex test the method in multidimensional flows where shocks coexist with shear, vorticity, material interfaces, wall effects, and three-dimensional compressible vortical dynamics. Comparisons with scalar IGR and WENO/TENO reference calculations are used to isolate the effect of the compression-directional tensor closure.

The remainder of the paper is organized as follows. Section~\ref{sec:method} defines the CoDeS regularized equations, the compression-directional stress closure, its structural properties, and the finite-volume discretization. Section~\ref{sec:result} presents the numerical results and comparisons with scalar IGR and high-order shock-capturing references. Section~\ref{sec:conclusion} summarizes the main findings and discusses limitations and future directions.

\section{Method} \label{sec:method}

\subsection{Overview}
This section defines the Compression-Directional Entropic Stress (CoDeS) method. The motivation and relation to information geometric regularization were summarized in Section~\ref{sec:intro}. Here we give the precise regularized equations, the tensor stress closure, its structural properties, and the finite-volume implementation.

The continuous CoDeS closure consists of three components. First, the Euler or Navier--Stokes equations are augmented by a conservative entropic-stress flux. Second, the stress tensor is constructed from the compressive part of the symmetric velocity-gradient tensor. Third, the scalar stress amplitude is obtained from a modified-Helmholtz equation. Together these components define the baseline CoDeS model used in the smooth-flow, contact, vortical-flow, shock-tube, multidimensional Riemann, viscous shock-tube, triple-point, and Taylor--Green vortex calculations reported below.

For completeness, we also describe two optional finite-volume safeguards. The first is a shock-localized face-normal compression source that augments only the scalar production term when a cell-centered strain approximation underresolves a severe one-sided compression. The second is a boundary-localized flux fallback that blends the high-order Euler flux with a low-order flux on a thin layer of faces adjacent to nonperiodic physical boundaries. These safeguards are implementation-level devices and are not part of the continuous CoDeS closure.

\subsection{Regularized Governing Equations}
Let
\begin{equation}
	U=(\rho, \rho \bm{u}, E)^T, \quad E=\rho e+\frac{1}{2} \rho|\bm{u}|^2,
\end{equation}
where $\rho$ is the density, $\bm{u}\in\mathbb{R}^d$ is the velocity vector, $E$ is the total energy per unit volume, and $e$ is the specific internal energy. For an ideal gas,
\begin{equation}
	p=(\gamma-1) \rho e .
\end{equation}

The inviscid CoDeS-regularized Euler equations are written as:
\begin{equation}
	\begin{aligned}
		&\partial_t \rho+\nabla \cdot(\rho \bm{u})=0 \\
		&\partial_t(\rho \bm{u})+\nabla \cdot\left(\rho \bm{u} \otimes \bm{u}+p \bm{I}+\bm{\Pi}_{\Sigma}\right)=0 \\
		&\partial_t E+\nabla \cdot\left[(E+p)\bm{u}+\bm{\Pi}_{\Sigma}\bm{u}\right]=0 .
	\end{aligned}
\end{equation}
where $\bm{\Pi}_{\Sigma}$ denotes the CoDeS entropic-stress tensor. The mass equation is unchanged. The momentum equation receives an additional conservative stress flux, and the energy equation receives the corresponding stress-work flux. The same tensor $\bm{\Pi}_{\Sigma}$ appears in both equations, which is essential for a consistent momentum--energy coupling.

For Navier--Stokes calculations, the physical viscous stress tensor $\boldsymbol{\tau}$ and the heat flux $\boldsymbol{q}$ are added separately,
\begin{equation}
	\begin{aligned}
		\partial_t(\rho \bm{u}) + \nabla\!\cdot\!\bigl( \rho \bm{u}\otimes \bm{u} + p\bm{I} + \boldsymbol{\Pi}_{\Sigma} - \boldsymbol{\tau} \bigr) &= 0,\\[4pt]
		\partial_t E + \nabla\!\cdot\!\bigl[ (E+p)\bm{u} + \boldsymbol{\Pi}_{\Sigma} \bm{u} - \boldsymbol{\tau} \bm{u} + \boldsymbol{q} \bigr] &= 0.
	\end{aligned}
\end{equation}
Unless otherwise stated, the viscous and thermal terms are modeled by the Newtonian stress and Fourier heat flux,
\begin{equation}
	\boldsymbol{\tau}
	=
	\mu\left[
	\nabla\bm{u}+(\nabla\bm{u})^T
	-\frac{2}{d}(\nabla\!\cdot\!\bm{u})\bm{I}
	\right],
	\qquad
	\boldsymbol{q}=-\kappa\nabla T ,
\end{equation}
CoDeS is therefore not a replacement for physical viscosity or thermal conduction. It is an additional entropic regularization stress intended to smooth shock-forming compressive regions.

In this work the stress tensor is written as
\begin{equation}
	\bm{\Pi}_{\Sigma}=\sigma \bm{M}.
\end{equation}
Here $\sigma(\boldsymbol{x}, t) \geq 0$ is a nonnegative scalar stress amplitude and $\boldsymbol{M}(\boldsymbol{x}, t) \succeq 0$ encodes the local compression direction. Rather than being a prescribed constant, $\sigma$ is obtained at each Runge-Kutta stage by solving a scalar modified-Helmholtz equation, and therefore varies in space and time with the evolving flow. Because the closure is built from a single scalar amplitude, only one such solve is required per Runge-Kutta stage.

\subsection{From Scalar IGR to Compression-Directional Stress}

The CoDeS closure is motivated by a structural limitation of the scalar multidimensional IGR source. In its multidimensional form, the scalar IGR production term contains
\begin{equation}
	Q_{\mathrm{IGR}} = \operatorname{tr}^2(\nabla \bm{u}) + \operatorname{tr}\!\left((\nabla \bm{u})^2\right).
\end{equation}
Introduce the decomposition
\begin{equation}
	\bm{A}=\nabla \bm{u}, \qquad  
	\bm{S}=\frac{1}{2}\left(\bm{A}+\bm{A}^T\right),  \qquad  
	\bm{\Omega}=\frac{1}{2}\left(\bm{A}-\bm{A}^T\right),  \qquad  
	\theta=\nabla \cdot \bm{u},
\end{equation}
where $\bm{S}$ is the symmetric strain-rate tensor, $\bm{\Omega}$ is the antisymmetric rotation tensor, and $\theta$ is the volumetric strain rate. Since
\begin{equation}
	\operatorname{tr}\!\left(\bm{A}^2\right) = \|\bm{S}\|_F^2 - \|\bm{\Omega}\|_F^2,
\end{equation}
the scalar source can be written as
\begin{equation}
	Q_{\mathrm{IGR}} = \theta^2 + \|\bm{S}\|_F^2 - \|\bm{\Omega}\|_F^2.
\end{equation}

The scalar multidimensional source therefore contains a negative contribution from rotation. In strongly vortical or shear-dominated regions, this term can render the scalar source sign-indefinite, even though the intended regularization mechanism is associated with compression. This observation motivates replacing the scalar multidimensional entropic pressure by a tensor stress constructed only from the compressive part of the symmetric strain-rate tensor.

The CoDeS closure is designed to satisfy four requirements. It should recover the one-dimensional compressive IGR source at planar shocks, vanish in pure rigid-body rotation, avoid density-gradient forcing at ideal contacts, and produce nonnegative formal entropy production in smooth compressible flows.

\subsection{Compression-Directional Entropic-Stress Closure}

We now define the CoDeS stress closure. The central idea is to separate the magnitude of the regularization from its direction. The scalar amplitude $\sigma$ determines how much entropic stress is applied, while the tensor $\boldsymbol{M}$ determines the local directions in which the stress acts. Unlike an isotropic artificial pressure, CoDeS aligns the stress with the compressive eigendirections of the local strain-rate tensor. This makes the regularization active in shock-forming compression, while suppressing it in expansion, rigid rotation, ideal contacts, and shear-dominated regions.

This construction is different from WENO/TENO-family reconstructions or classical nonlinear limiters, which stabilize shocks primarily by nonlinear adaptation of the discrete reconstruction. CoDeS instead supplies shock
regularization through a conservative PDE-level stress flux aligned with local compression. This separation allows one to use high-order reconstruction while reducing the reliance on reconstruction-level nonlinear limiting in smooth compressive, shear, contact, and vortical regions.

Let
\begin{equation}
	\boldsymbol{L} = \nabla\boldsymbol{u},\qquad
	\boldsymbol{S} = \frac12(\boldsymbol{L}+\boldsymbol{L}^T),\qquad
	\theta = \operatorname{tr}\boldsymbol{S} = \nabla\cdot\boldsymbol{u}.
	\label{eq:codes-kinematics}
\end{equation}
Here $\boldsymbol{S}$ is the symmetric strain-rate tensor and $\theta$ is the volumetric strain rate. Negative $\theta$ corresponds to local volume compression, whereas positive $\theta$ corresponds to local expansion.

Using the positive-part notation
\begin{equation}
	z_{+}=\max(z,0),
	\label{eq:positive-part}
\end{equation}
we define the scalar net-compression rate
\begin{equation}
	c=(-\theta)_{+}.
	\label{eq:net-compression}
\end{equation}
Thus $c=-\theta$ in locally converging flow and $c=0$ in locally expanding or volume-preserving flow. This scalar gate ensures that CoDeS is triggered by net compression rather than by arbitrary shear or expansion.

To determine the directions of compression, let the spectral decomposition of $\boldsymbol{S}$ be
\begin{equation}
	\boldsymbol{S} = \sum_{a=1}^{d} \lambda_a \, \boldsymbol{r}_a \boldsymbol{r}_a^T ,
	\label{eq:S-spectral}
\end{equation}
where $\lambda_a$ are the principal strain rates and $\boldsymbol{r}_a$ are the corresponding orthonormal eigenvectors. A principal direction is compressive when $\lambda_a<0$. We therefore define the compression tensor
\begin{equation}
	\boldsymbol{N} = \sum_{a=1}^{d} (-\lambda_a)_+ \, \boldsymbol{r}_a \boldsymbol{r}_a^T \succeq 0,\qquad \operatorname{tr}\boldsymbol{N} = \sum_{a=1}^{d} (-\lambda_a)_+.
	\label{eq:compression-tensor}
\end{equation}
In \eqref{eq:compression-tensor}, the notation $(-\lambda_a)_+$ keeps only compressive principal strain rates: if $\lambda_a<0$, then $(-\lambda_a)_+=-\lambda_a>0$, while if $\lambda_a\geq0$, then $(-\lambda_a)_+=0$. Hence $\boldsymbol{N}$ records both the directions and magnitudes of local principal compression, while $\operatorname{tr} \boldsymbol{N}$ measures the total principal compression. In pure expansion or rigid-body rotation, $\boldsymbol{N}=0$.

The scalar CoDeS production source is then defined as
\begin{equation}
	b = \beta_d \, c \, \operatorname{tr}\boldsymbol{N} = \beta_d \, (-\theta)_+ \, \operatorname{tr}\boldsymbol{N}.
	\label{eq:codes-source}
\end{equation}
The product structure in \eqref{eq:codes-source} requires both net volumetric compression and compressive principal strain. Consequently, $b$ is nonnegative by construction and vanishes in expansion, rigid rotation, and ideal contacts with constant velocity.

The dimension-dependent coefficient $\beta_d$ is chosen so that CoDeS agrees with the vorticity-free scalar IGR source in isotropic compression. Here $d$ denotes the spatial dimension. Consider
\begin{equation}
	\boldsymbol{S} = -a\boldsymbol{I},\quad a>0.
	\label{eq:isotropic-compression}
\end{equation}

Substituting \eqref{eq:isotropic-compression} into \eqref{eq:codes-kinematics}, \eqref{eq:net-compression}, and \eqref{eq:compression-tensor} gives
\begin{equation}
	\theta = -da,\quad c=da,\quad \boldsymbol{N}=a\boldsymbol{I}, \quad \operatorname{tr}\boldsymbol{N}=da,
	\label{eq:isotropic-compression-values}
\end{equation}
Therefore the CoDeS source becomes
\begin{equation}
	b = \beta_d d^2 a^2.
	\label{eq:isotropic-codes-source}
\end{equation}
The corresponding vorticity-free scalar multidimensional IGR source is
\begin{equation}
	\theta^2+\|\boldsymbol{S}\|_F^2 .
	\label{eq:scalar-igr-source}
\end{equation}
For $\boldsymbol{S} = -a \boldsymbol{I}$, this evaluates to
\begin{equation}
	\theta^2+\|\boldsymbol{S}\|_F^2 = d^2 a^2+d a^2=d(d+1) a^2 .
	\label{eq:isotropic-igr-source}
\end{equation}
Equating \eqref{eq:isotropic-codes-source} and \eqref{eq:isotropic-igr-source} gives
\begin{equation}
	\beta_d d^2 a^2=d(d+1) a^2,
\end{equation}
and hence
\begin{equation}
	\beta_d=\frac{d+1}{d} .
	\label{eq:beta-d}
\end{equation}

The compression-direction tensor is obtained by normalizing $\boldsymbol{N}$,
\begin{equation}
	\boldsymbol{M} = \frac{\boldsymbol{N}}{\operatorname{tr}\boldsymbol{N} + \varepsilon_M},
	\label{eq:compression-direction-tensor}
\end{equation}
where $\varepsilon_M>0$ prevents division by zero in regions without compression. When $\operatorname{tr}\boldsymbol{N}$ is nonzero and $\varepsilon_M$ is negligible, $\boldsymbol{M}$ has approximately unit trace and preserves only the directional distribution of the local compression.

The scalar amplitude $\sigma=\sigma(\boldsymbol{x},t)$ is obtained from the modified-Helmholtz equation
\cite{CaoSchaefer2023BarotropicIGR,CaoSchaefer2024PressurelessIGR}
\begin{equation}
	\rho^{-1}\sigma - \nabla\!\cdot\!\bigl(\alpha\rho^{-1}\nabla\sigma\bigr) = \alpha b .
	\label{eq:sigma-helmholtz}
\end{equation}
Thus $\sigma$ is not a prescribed constant. It is recomputed at each Runge--Kutta stage and varies in space and time with the evolving compression field. Only this scalar equation is solved globally, the tensor direction
$\boldsymbol{M}$ is computed locally from the strain-rate eigenstructure.

Finally, the CoDeS stress tensor is
\begin{equation}
	\boldsymbol{\Pi}_{\Sigma} = \sigma \boldsymbol{M}.
	\label{eq:codes-stress}
\end{equation}
This form gives a tensor-valued, compression-aligned stress while requiring only one scalar modified-Helmholtz solve per Runge--Kutta stage.

The parameter $\alpha$ controls the regularization length scale. In the present finite-volume implementation, following \cite{RadhakrishnanEtAl2026FVMIGR}, we set
\begin{equation}
	\alpha=C_\alpha h^2,
	\label{eq:alpha-scaling}
\end{equation}
where $h$ is a representative local mesh spacing. The modified-Helmholtz solve therefore smooths $\sigma$ over a length scale of order $\sqrt{\alpha}$. In practice, a positivity guard $\sigma\leftarrow\max(\sigma,0)$ can be applied after the solve to remove small negative values caused by discrete solver error.

\subsection{Structural Properties of the CoDeS Closure}
\subsubsection{Recovery of One-dimensional Compressive IGR}
In one spatial dimension,
\begin{equation}
	S=u_x,\quad \theta=u_x,\quad N=(-u_x)_+ .
\end{equation}
For compression, $u_x<0$, and therefore
\begin{equation}
	c=-u_x, \quad \operatorname{tr} N=-u_x, \quad \beta_1=2 .
\end{equation}

The CoDeS source becomes
\begin{equation}
	b=2 u_x^2 .
\end{equation}
Moreover, $M \approx 1$, so $\boldsymbol{\Pi}_{\Sigma}=\sigma$. The modified-Helmholtz equation reduces to
\begin{equation}
	\rho^{-1} \sigma-\partial_x\left(\alpha \rho^{-1} \partial_x \sigma\right)=2 \alpha u_x^2, \quad u_x<0 .
\end{equation}

Thus the CoDeS closure recovers the compressive part of the one-dimensional scalar IGR entropic-pressure equation. The key modeling difference is that CoDeS is deliberately not activated in expansion.

\subsubsection{Rigid-Body Rotation}

For pure rigid-body rotation, the velocity-gradient tensor is skew-symmetric:
\begin{equation}
	\boldsymbol{L}^T=-\boldsymbol{L}.
\end{equation}
Therefore, by the definition of the symmetric strain-rate tensor in
\eqref{eq:codes-kinematics},
\begin{equation}
	\boldsymbol{S} = \frac12(\boldsymbol{L}+\boldsymbol{L}^T) = 0,
	\qquad
	\theta=\operatorname{tr}\boldsymbol{S}=0 .
\end{equation}
Using the definitions of the compression rate, compression tensor, and source
term in \eqref{eq:net-compression}, \eqref{eq:compression-tensor}, and
\eqref{eq:codes-source}, we obtain
\begin{equation}
	c=0, \qquad \boldsymbol{N}=0, \qquad b=0 .
\end{equation}
It then follows from \eqref{eq:compression-direction-tensor} that
\begin{equation}
	\boldsymbol{M}=0,
\end{equation}
and hence, by the stress definition \eqref{eq:codes-stress},
\begin{equation}
	\boldsymbol{\Pi}_{\Sigma}=0 .
\end{equation}
Thus CoDeS generates no entropic stress from pure rigid-body rotation. This removes the sign-indefinite vorticity contribution present in the raw scalar multidimensional IGR source.

\subsubsection{Contact Compatibility}

An ideal stationary contact discontinuity has constant pressure and velocity, with a density or entropy jump. In such a state, the velocity gradient vanishes:
\begin{equation}
	\nabla\boldsymbol{u}=0 .
\end{equation}
Therefore, by \eqref{eq:codes-kinematics},
\begin{equation}
	\boldsymbol{S}=0, \qquad \theta=0 .
\end{equation}
Using \eqref{eq:net-compression}, \eqref{eq:compression-tensor}, and \eqref{eq:codes-source}, we obtain
\begin{equation}
	c=0, \qquad \boldsymbol{N}=0, \qquad b=0 .
\end{equation}
Consequently, \eqref{eq:compression-direction-tensor} gives
\begin{equation}
	\boldsymbol{M}=0,
\end{equation}
and the CoDeS stress vanishes by \eqref{eq:codes-stress}:
\begin{equation}
	\boldsymbol{\Pi}_{\Sigma}=0 .
\end{equation}
CoDeS therefore introduces no density-gradient, entropy-gradient, or pressure-gradient forcing at an ideal contact. The closure responds to compressive velocity gradients, not to thermodynamic jumps alone.

\subsubsection{Formal Entropy-production Mechanism}\label{sec:entropy-production}
For smooth inviscid flows with an added conservative stress flux, the thermodynamic identity formally gives
\begin{equation}
	\rho T\frac{Ds}{Dt} = -\boldsymbol{\Pi}_{\Sigma}:\boldsymbol{S} .
\end{equation}
For CoDeS,
\begin{equation}
	\boldsymbol{\Pi}_{\Sigma}=\sigma \boldsymbol{M}, \quad \sigma \geq 0, \quad \boldsymbol{M} \succeq 0 .
\end{equation}
Since $\boldsymbol{M}$ is supported only in the compressive eigenspace of $\boldsymbol{S}$,
\begin{equation}
	\boldsymbol{M} : \boldsymbol{S} \leq 0,
\end{equation}
and therefore
\begin{equation}
	-\boldsymbol{\Pi}_{\Sigma}:\boldsymbol{S} \geq 0.
\end{equation}

This yields a clean continuous entropy-production mechanism in smooth compressible flows. The property is used here as a design principle for the method, and is not claimed as a complete discrete entropy-stability theorem or as a convergence proof to entropy solutions of the unregularized Euler equations.

\subsection{Finite-Volume Discretization}
Let the computational domain be partitioned into control volumes $K_i$ of measure $|K_i|$, with cell averages
\begin{equation}
	\overline{\boldsymbol{U}}_i(t)= \frac{1}{|K_i|} \int_{K_i} \boldsymbol{U}(\boldsymbol{x},t)\,d\boldsymbol{x}.
\end{equation}
The finite-volume semi-discretization has the form
\begin{equation}
	\frac{d\overline{\boldsymbol{U}}_i}{dt} = -\frac{1}{|K_i|} \sum_{f\subset\partial K_i} |f| \, \widehat{\boldsymbol{F}}_{i,f},
\end{equation}
and the numerical flux is decomposed as
\begin{equation}
	\widehat{\boldsymbol{F}}_{i,f} = \widehat{\boldsymbol{F}}_{\mathrm{E},i,f} + \widehat{\boldsymbol{F}}_{\Sigma,i,f},
\end{equation}
where $\widehat{\boldsymbol{F}}_{\mathrm{E},i,f}$ is the baseline Euler or Navier--Stokes flux and $\widehat{\boldsymbol{F}}_{\Sigma,i,f}$ is the CoDeS stress flux. The baseline flux may be obtained from any standard finite-volume reconstruction and Riemann solver. The CoDeS regularization supplies an additional conservative stress flux.

For an interior face $f=i \mid j$ with unit normal $n_{f}$ pointing from cell $i$ to cell $j$, the CoDeS mass flux vanishes,
\begin{equation}
	\widehat{\boldsymbol{F}}_{\Sigma,f}^{\rho} = 0 ,
\end{equation}
and the momentum contribution is
\begin{equation}
	\widehat{\boldsymbol{F}}_{\Sigma,f}^{\rho\boldsymbol{u}} = \widehat{\boldsymbol{\Pi}}_{\Sigma,f}\,\boldsymbol{n}_f .
\end{equation}
A simple conservative central choice is
\begin{equation}
	\widehat{\boldsymbol{\Pi}}_{\Sigma,f} = \frac{1}{2}\bigl( \boldsymbol{\Pi}_{\Sigma}^{-} + \boldsymbol{\Pi}_{\Sigma}^{+} \bigr),
\end{equation}
where $-$ and $+$ denote the two sides of the face.

The energy contribution is the corresponding stress-work flux. In the present implementation we use the side-averaged stress work,
\begin{equation}
	\widehat{F}_{\Sigma,f}^{E} = \frac{1}{2}\Bigl[ \bigl(\boldsymbol{\Pi}_{\Sigma}^{-}\boldsymbol{n}_f\bigr)\!\cdot\!\boldsymbol{u}^{-} + \bigl(\boldsymbol{\Pi}_{\Sigma}^{+}\boldsymbol{n}_f\bigr)\!\cdot\!\boldsymbol{u}^{+}\Bigr].
\end{equation}
The same face flux, with opposite normal directions, is used for the neighboring control volume, so the CoDeS contribution remains conservative.

The scalar modified-Helmholtz equation for $\sigma$ is discretized on the same finite-volume mesh. A representative cell-centered finite-volume discretization is
\begin{equation}
	\frac{\sigma_i}{\rho_i} - \frac{1}{|K_i|} \sum_{f\subset\partial K_i} |f| \left(\alpha\rho^{-1}\right)_f
	\frac{\sigma_j-\sigma_i}{d_f} = \alpha_i b_i .
\end{equation}
Here $|K_i|$ is the measure of cell $K_i$, $|f|$ is the measure of face $f$, and $j$ denotes the cell adjacent to $i$ across face $f$. The quantity $d_f$ is the distance between the two cell centers measured in the face-normal direction. Thus $(\sigma_j-\sigma_i)/d_f$ is a two-point approximation of the normal derivative $\nabla\sigma\cdot\boldsymbol{n}_f$ at face $f$; in one spatial dimension it reduces to the usual centered difference between adjacent cells. The coefficient $\left(\alpha\rho^{-1}\right)_f$ is a consistent face average.

Periodic or homogeneous Neumann conditions are used for $\sigma$ unless the test problem requires otherwise. Because the operator contains a positive mass term, the resulting linear system is better conditioned than a pure Poisson problem and can be solved with standard iterative elliptic solvers.

\subsection{Wall Treatment for Stress Tensor}
At reflecting or slip walls, stress tensors require a special projection. For a wall with outward unit normal $\boldsymbol{n}$, the no-penetration condition gives $\boldsymbol{u}\!\cdot\!\boldsymbol{n}=0$. For a scalar stress, the wall stress work vanishes automatically. For a stress tensor, however, $\boldsymbol{\Pi}_{\Sigma}\boldsymbol{n}$ may contain tangential components, which would introduce an artificial tangential wall traction and nonzero wall stress work. To avoid this, we use a projected wall stress flux.

At reflecting or slip walls, the CoDeS wall flux is
\begin{equation}
	\begin{aligned}
		\widehat{F}_{\Sigma,\mathrm{wall}}^{\rho}&=0,\\[4pt]
		\widehat{\boldsymbol{F}}_{\Sigma,\mathrm{wall}}^{\rho\boldsymbol{u}}
		&= \bigl(\boldsymbol{n}^{T}\boldsymbol{\Pi}_{\Sigma}\boldsymbol{n}\bigr)\boldsymbol{n}, \\[4pt]
		\widehat{F}_{\Sigma,\mathrm{wall}}^{E}&=0 .
	\end{aligned}
\end{equation}
Only the normal--normal component of the stress tensor contributes to wall-normal momentum, and the CoDeS stress performs no work at an impermeable slip wall.

\subsection{Shock-Localized Face-Normal Compression Source}
The baseline CoDeS source is computed from a cell-centered approximation of the symmetric velocity gradient. This construction yields the desired continuous structure. In rare pathological configurations, however, the cell-centered strain may underrepresent the one-sided compression seen by the numerical Riemann problems at cell faces. For such configurations we provide an optional discrete face-normal compression source as a robustness extension of the one-channel CoDeS model. This term is not a separate physical model and does not introduce a second stress tensor. It augments only the scalar source entering the same modified-Helmholtz equation for $\sigma$.

Let $i$ denote a cell of a Cartesian finite-volume grid, and let $\boldsymbol{e}_m$ be the coordinate direction associated with velocity component $u_m$. The face-normal source is a discrete, face-normal compression sensor motivated by compression-based artificial-viscosity shock regularization \cite{one-sided-coordinate-normal-compression1, one-sided-coordinate-normal-compression2}. The one-sided coordinate-normal compression rates  are defined as
\begin{equation}
	c_{m,-,i} = \max\left( - \frac{u_{m,i}-u_{m,i-\boldsymbol{e}_m}}{\Delta x_{m,i}}, 0 \right),
	\qquad
	c_{m,+,i} = \max\left( - \frac{u_{m,i+\boldsymbol{e}_m}-u_{m,i}}{\Delta x_{m,i}}, 0 \right).
\end{equation}

Only converging one-sided face-normal velocity differences contributions. The raw face source is the maximum squared one-sided compression,
\begin{equation}
	b_{f,i} = \max_{m=1,\ldots,d} \left( c_{m,-,i}^{2}, c_{m,+,i}^{2} \right),
\end{equation}
so that $b_{f,i}$ has the same physical dimension as the baseline strain source.

To keep the added source localized to shocks, it is multiplied by a smooth pressure-jump activation. For cell $i$, let $\mathcal{N}(i)$ be the set of face-neighboring cells. We define
\begin{equation}
	\eta_i 	= \max_{j\in\mathcal{N}(i)} \frac{|p_j-p_i|} {|p_j|+|p_i|+\epsilon_p},
\end{equation}
where $\epsilon_p$ is a small pressure scale used only as a numerical guard. The activation weight is
\begin{equation}
	w_{\eta,i} = S_{01} \left( \frac{\eta_i-\eta_0}{\eta_1-\eta_0} \right),
\end{equation}
with the clamped cubic smoothstep
\begin{equation}
	S_{01}(z)= 	\begin{cases}
		0, & z\le 0,\\
		z^2(3-2z), & 0<z<1,\\
		1, & z\ge 1 .
	\end{cases}
\end{equation}
The constants $\eta_0$ and $\eta_1$ are the lower and upper pressure-jump thresholds for activation.

The face-normal contribution is then
\begin{equation}
	b_{\mathrm{face},i} = \kappa_f\,w_{\eta,i}\,b_{f,i},
\end{equation}
where $\kappa_f\ge 0$ is a dimensionless coefficient. Setting $\kappa_f=0$ disables this safeguard and recovers the baseline source. When the safeguard is enabled, the scalar source used in the modified-Helmholtz equation is
\begin{equation}
	b_i^{*} = \max\left( b_i,\, b_{\mathrm{face},i} \right),
	\qquad
	b_i=\beta_d(-\theta_i)_+\operatorname{tr}\boldsymbol{N}_i ,
\end{equation}
and the modified-Helmholtz problem is solved as
\begin{equation}
	\rho_i^{-1}\sigma_i-\nabla \cdot\left(\alpha_i\rho_i^{-1}\nabla\sigma_i\right) = \alpha_i b_i^{*}.
\end{equation}

In the face-source variant used in this work, the tensor direction is not replaced by a face-normal tensor. The stress direction remains
\begin{equation}
	\boldsymbol{M}_i = \frac{\boldsymbol{N}_i}{\operatorname{tr}\boldsymbol{N}_i+\varepsilon_M}.
\end{equation}

The face source therefore increases the smoothed stress amplitude in shock cells, while the stress direction remains determined by the compressive eigenspaces of the symmetric strain tensor. Since the augmented source remains nonnegative and $\boldsymbol{M}$ is unchanged, the formal entropy-production argument of Section~\ref{sec:entropy-production} is not altered at the continuous design level.

This face source is distinct from the boundary flux fallback described below. It does not compute a Harten--Lax--van Leer with Einfeldt waves (HLLE) flux, does not blend the Euler numerical flux. Instead, it is an interior finite-volume shock sensor that modifies only the scalar source in the CoDeS stress-amplitude solver, and should be interpreted as an implementation-level safeguard for unusually harsh shock and boundary interactions.

\subsection{Boundary-Localized Flux Fallback}
The CoDeS stress regularizes shock-forming compression in the domain interior. In rare pathological configurations, the physical boundary condition and its ghost-state construction can locally violate the idealized regularized characteristic structure. In our experience, such failures occur when a strong shock develops directly at a nonperiodic physical boundary, even though comparably strong interior shocks remain stable. Increasing the CoDeS amplitude alone is not an efficient cure for these boundary-localized failures. We therefore apply a localized fallback to the hyperbolic Euler flux in a thin layer of faces adjacent to nonperiodic physical boundaries.

This fallback is not a modification of the CoDeS closure. It does not change the CoDeS source (including the face-source augmentation when that is enabled), the direction $\boldsymbol{M}$, the amplitude $\sigma$, the modified-Helmholtz solve, or the CoDeS stress flux. It acts only on selected Euler numerical fluxes in a prescribed boundary layer.

Let $\boldsymbol{F}_f^{\mathrm{high}}$ denote the high-order reconstructed Euler flux at face $f$. Outside a prescribed layer adjacent to a nonperiodic physical boundary, we set $\boldsymbol{F}_f^{\mathrm{E}} = \boldsymbol{F}_f^{\mathrm{high}}$. On eligible boundary-layer faces, we compute a first-order HLLE flux $\boldsymbol{F}_f^{\mathrm{HLLE}}$\cite{HLL, HLLE} from the face-adjacent cell or ghost states and replace the Euler flux by the convex blend
\begin{equation}
	\boldsymbol{F}_f^{\mathrm{E}} = (1-\omega_f)\boldsymbol{F}_f^{\mathrm{high}} + \omega_f \boldsymbol{F}_f^{\mathrm{HLLE}},\qquad 0\le \omega_f\le 1 .
\end{equation}
The same blended face flux is used in the residuals of the cells sharing the face, so the finite-volume update remains conservative.

The fallback is considered only within a prescribed number $N_{b}$ of face layers adjacent to a nonperiodic physical boundary. In practice, one layer is sufficient. Periodic boundaries are excluded, and the fallback therefore remains inactive in the domain interior independently of shock strength.

For an eligible face, let the primitive states associated with the two adjacent cell or ghost states be
\begin{equation}  
	\left(\rho_L, u_L, v_L, p_L, a_L\right), \quad\left(\rho_R, u_R, v_R, p_R, a_R\right),  
\end{equation}
where $a$ is the sound speed. Let $\bm{n}=\left(n_x, n_y\right)$ be the face normal. The normal velocities are
\begin{equation}  
	u_{n, L}=u_L n_x+v_L n_y, \quad u_{n, R}=u_R n_x+v_R n_y .  
\end{equation}
The pressure-ratio sensor is
\begin{equation}  
	R_p=\frac{\max \left(\left|p_L\right|,\left|p_R\right|\right)}{\max \left(\min \left(\left|p_L\right|,\left|p_R\right|\right), \epsilon_p\right)} .  
\end{equation}
For physical positive-pressure states, this reduces to the usual pressure ratio. The absolute values and the small positive number $\epsilon_p$ are numerical guards.

The pressure contribution to the fallback weight is
\begin{equation}  
	\omega_p=S_{01}\left(\frac{\log \left(\max \left(R_p, 1\right)\right)-\log \left(R_{p, 0}\right)}{\log \left(R_{p, 1}\right)-\log \left(R_{p, 0}\right)}\right),  
\end{equation}
where $R_{p, 0}$ and $R_{p, 1}$ are lower and upper pressure-ratio thresholds, and $S_{01}$ is the clamped cubic smoothstep
\begin{equation}  
	S_{01}(z)= \begin{cases}0, & z \leq 0, \\ z^2(3-2 z), & 0<z<1, \\ 1, & z \geq 1 .\end{cases}  
\end{equation}
An optional normal-velocity-jump Mach gate is defined by
\begin{equation}  
	M_c=\frac{\left|u_{n, R}-u_{n, L}\right|}{\max \left(\frac{1}{2}\left(a_L+a_R\right), \epsilon_p\right)} .  
\end{equation}
When $M_{c, 1}>M_{c, 0}$, the corresponding weight is
\begin{equation}  
	\omega_c=S_{01}\left(\frac{M_c-M_{c, 0}}{M_{c, 1}-M_{c, 0}}\right) .
\end{equation}
When $M_{c, 1} \leq M_{c, 0}$, this gate is disabled and $\omega_{c}=1$. The final fallback weight is
\begin{equation}  
	\omega_f=\min \left\{\max \left(\omega_{\max } \omega_p \omega_c, 0\right), 1\right\} .
\end{equation}
Thus $\omega_{f}=0$ recovers the original high-order Euler flux, while $\omega_{f}=1$ gives a pure first-order flux on that face. The HLLE flux is chosen as the first-order flux in this study.

The fallback is deliberately localized. It is used only to regularize local shock-boundary Riemann problems and ghost-state incompatibilities in severe bounded-domain tests. Because the limiting flux is first-order HLLE, activation of the fallback locally reduces the formal order of accuracy near the affected boundary faces.

\section{Results and Discussion} \label{sec:result}

This section evaluates the CoDeS closure on a hierarchy of test problems ranging from diagnostic one-dimensional cases to multidimensional shock, shear, wall-bounded, and multi-material flows. Unless otherwise stated, the gas is ideal with $\gamma=1.4$, and all reported CoDeS calculations use the baseline closure of Section~\ref{sec:method}. The stress amplitude is obtained from the modified-Helmholtz equation, and the stress tensor is aligned with the local compressive eigenspace. The shock-localized face-normal source and the boundary-localized flux fallback are disabled in every test except the Mach-3 slot jet, where they are explicitly identified as robustness safeguards. No additional artificial-viscosity term, nonlinear limiter, or WENO-type nonlinear reconstruction is employed in the CoDeS runs.

In all comparisons between CoDeS and scalar IGR, the discretization is held fixed except for the entropic-stress closure. The same cell-centered finite-volume framework, local Lax--Friedrichs/Rusanov (LF/Rusanov) convective flux \cite{rusanov, lax}, time integrator, and reconstruction order are used in both cases. When a WENO reference is included, the linear reconstruction used with CoDeS is chosen to match the formal order of the WENO reconstruction. For example, CoDeS uses a fifth-order reconstruction in comparison with WENO-5 and seventh-order reconstruction in comparisons with WENO-7. This protocol isolates the effect of the compression-directional closure and demonstrates that CoDeS is not tied to a particular reconstruction stencil.

For tests initialized from discontinuous Riemann data, the numerical initial condition is obtained by smoothing the primitive variables across the discontinuity,
\begin{equation}
	q(x,0)=(1-S)q_L+S q_R,\qquad
	S(x)=\frac{1}{2}\left[1+\tanh\left(\frac{x-x_0}{\delta}\right)\right],
\end{equation}
where $q=(\rho,u,p)$ and the smoothing width $\delta$ is specified for each test. The plotted reference solution is the exact sharp-interface Euler Riemann solution unless otherwise noted.

\begin{figure}[t]
	\centering
	\begin{subfigure}[t]{0.48\textwidth}
		\centering
		\includegraphics[width=\linewidth]{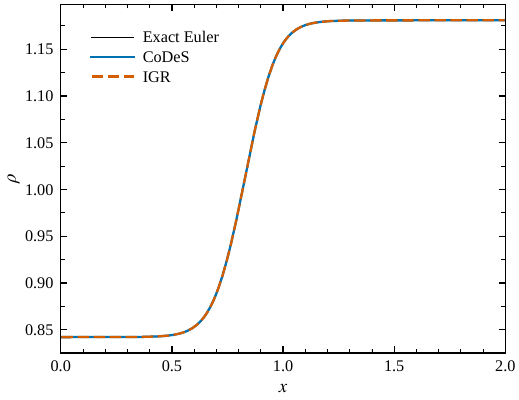}
		\caption{Density $\rho$.}
		\label{fig:smooth-simple-wave-rho}
	\end{subfigure}
	\hfill
	\begin{subfigure}[t]{0.48\textwidth}
		\centering
		\includegraphics[width=\linewidth]{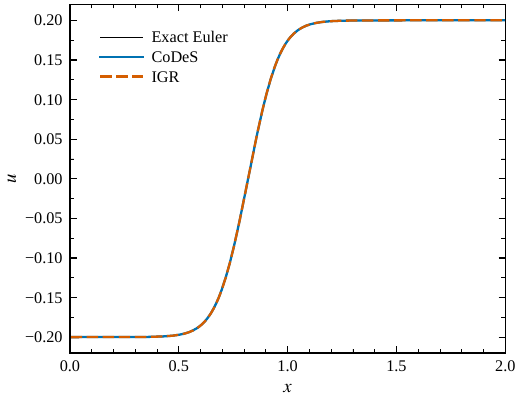}
		\caption{Velocity $u$.}
		\label{fig:smooth-simple-wave-u}
	\end{subfigure}
	
	\vspace{0.5em}
	
	\begin{subfigure}[t]{0.48\textwidth}
		\centering
		\includegraphics[width=\linewidth]{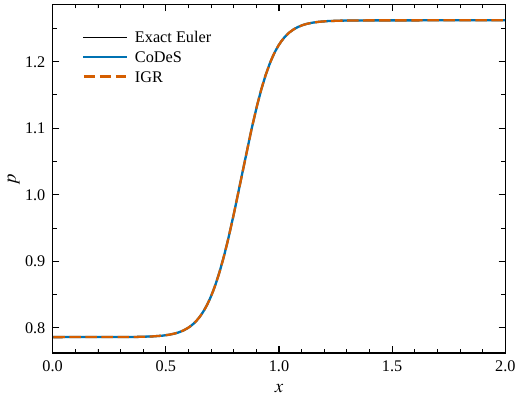}
		\caption{Pressure $p$.}
		\label{fig:smooth-simple-wave-p}
	\end{subfigure}
	\hfill
	\begin{subfigure}[t]{0.48\textwidth}
		\centering
		\includegraphics[width=\linewidth]{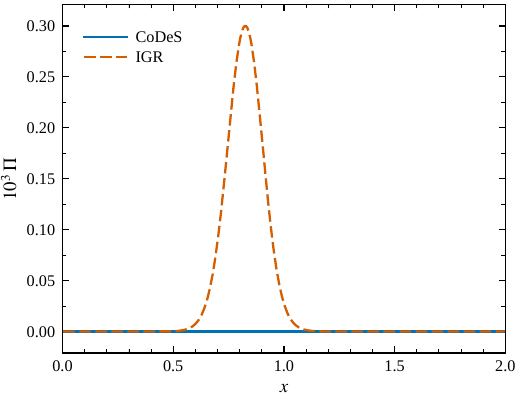}
		\caption{Entropic stress $\Pi$.}
		\label{fig:smooth-simple-wave-pi}
	\end{subfigure}
	
	\caption{
		Smooth isentropic simple-wave expansion at $t=0.1$ on the $n=320$ grid.
		The CoDeS solution remains visually indistinguishable from the exact Euler solution in
		$\rho$, $u$, and $p$, while the CoDeS entropic stress remains zero to roundoff.
		In contrast, scalar IGR generates a finite stress inside the smooth expansion fan.
	}
	\label{fig:smooth-simple-wave-profiles}
\end{figure}
\begin{figure}[t]
	\centering
	\includegraphics[width=0.68\textwidth]{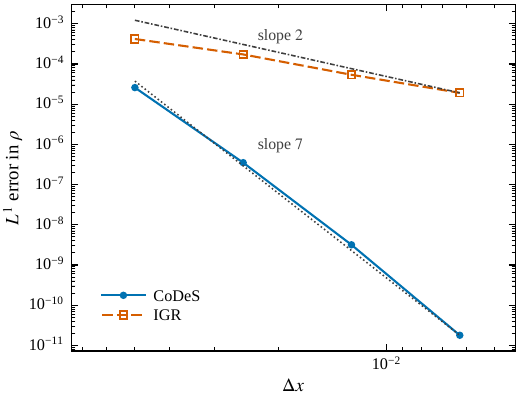}
	\caption{
		$L^1$ density-error convergence for the smooth isentropic simple-wave expansion.
		CoDeS recovers the high-order behavior of the underlying seventh-order finite-volume
		discretization, whereas scalar IGR converges at a substantially lower rate because it
		activates a finite stress in the smooth expansion region.
	}
	\label{fig:smooth-simple-wave-convergence}
\end{figure}

\subsection{Smooth Isentropic Simple-wave Expansion}
We first consider a smooth right-going isentropic simple wave prior to wave breaking. The case is deliberately non-shocking, since the wave is expansive throughout the diagnostic interval and a shock regularization should therefore remain inactive. This setup directly tests whether the closure distinguishes shock-forming compression from smooth expansion while preserving the accuracy of the underlying high-order finite-volume discretization.

The domain is $x\in[0,2]$ with zero-gradient extrapolation boundary conditions. The initial velocity is

\begin{equation}
	u(x,0)=(1-S)u_L+S u_R,\qquad
	S(x)=\frac{1}{2}\left[1+\tanh\left(\frac{x-x_c}{w}\right)\right],
\end{equation}
with $u_L=-0.2$, $u_R=0.2$, $x_c=0.7$, and $w=0.12$. Density and pressure follow from the right-going simple-wave invariant,
\begin{equation}
	u-\frac{2a}{\gamma-1}
	=u_{\mathrm{ref}}-\frac{2a_{\mathrm{ref}}}{\gamma-1},\qquad
	a_{\mathrm{ref}}=\sqrt{\gamma p_{\mathrm{ref}}/\rho_{\mathrm{ref}}},
\end{equation}
where $(\rho_{\mathrm{ref}},p_{\mathrm{ref}},u_{\mathrm{ref}})=(1,1,0)$. Hence
\begin{equation}
	a=a_{\mathrm{ref}}+\frac{\gamma-1}{2}u,\qquad p=K\rho^\gamma,\qquad K=1 .
\end{equation}
Since $u$ increases monotonically through the transition, the right-going characteristic speed
\begin{equation}
	u+a=a_{\mathrm{ref}}+\frac{\gamma+1}{2}u
\end{equation}
also increases. The characteristics therefore diverge rather than collide, and the solution remains smooth throughout the time interval considered.

The exact Euler solution is obtained by characteristic tracing. For a point $x$ at time $t$, the footpoint $\xi$ satisfies
\begin{equation}
	x=\xi+\{u(\xi,0)+a(\xi,0)\}t .
\end{equation}
Initial and final reference cell averages are evaluated using Gauss--Legendre quadrature. The convergence study uses $n=40,80,160,320$ cells and final time $t=0.1$. To ensure that the reported errors are dominated by the spatial discretization, the classical fourth-order Runge--Kutta (RK4) scheme is used with
\begin{equation}
	\Delta t=\mathrm{CFL}\frac{h^2}{\max(|u|+a)},\qquad \mathrm{CFL}=1 .
\end{equation}
Errors are measured in the expansion window $x\in[0.58,0.82]$, where $u_x>0$ and the difference between CoDeS and scalar IGR is most clearly exposed.

Figure~\ref{fig:smooth-simple-wave-profiles} shows density, velocity, pressure, and the entropic stress on the $n=320$ grid, together with the density-error convergence. The CoDeS solution is visually indistinguishable from the exact Euler solution in all primitive variables, and the CoDeS stress remains zero to roundoff throughout the smooth expansion fan. This is the intended one-dimensional behavior. Because the CoDeS source is gated by compression, an expansive region with $u_x>0$ generates no regularizing stress.

In contrast, scalar IGR produces a finite stress even within the same smooth expansion fan. This response is not a shock regularization, but rather a false positive caused by expansion. The issue is also reflected in the sign of the stress-work term: in one-dimensional smooth flow, the formal contribution is proportional to $-\pi u_x$. Thus, when $u_x>0$ in an expansive region, a positive scalar stress $\pi$ gives a contribution with the opposite sign from entropy production.

The effect of this spurious activation is summarized in Figure~\ref{fig:smooth-simple-wave-convergence}, which shows the $L^1$ density error as the grid is refined. CoDeS follows the high-order behavior of the underlying seventh-order finite-volume discretization, whereas scalar IGR converges at a substantially lower rate. The corresponding errors and observed rates for all three primitive variables are reported in Table~\ref{tab:smooth-simple-wave-convergence}. The figure provides a compact visual summary of the convergence behavior, while the table gives the quantitative data for $\rho$, $u$, and $p$.

\begin{table}[h]
	\centering
	\small
	\caption{
		$L^1$ errors and observed convergence rates for the smooth isentropic simple-wave
		expansion at $t=0.1$. Errors are measured in the expansion window
		$x\in[0.58,0.82]$.
	}
	\label{tab:smooth-simple-wave-convergence}
	\setlength{\tabcolsep}{4pt}
	\begin{tabular}{@{}l r c c c c c c@{}}
		\toprule
		Model & $n$ & $L^1_\rho$ & Rate & $L^1_u$ & Rate & $L^1_p$ & Rate \\
		\midrule
		CoDeS &  40 & $2.5829\times10^{-5}$  & --   & $3.4211\times10^{-5}$  & --   & $4.3503\times10^{-5}$  & --   \\
		CoDeS &  80 & $3.5044\times10^{-7}$  & 6.20 & $4.8741\times10^{-7}$  & 6.13 & $6.1667\times10^{-7}$  & 6.14 \\
		CoDeS & 160 & $3.1740\times10^{-9}$  & 6.79 & $4.5903\times10^{-9}$  & 6.73 & $5.7015\times10^{-9}$  & 6.76 \\
		CoDeS & 320 & $1.7837\times10^{-11}$ & 7.48 & $2.0201\times10^{-11}$ & 7.83 & $2.0775\times10^{-11}$ & 8.10 \\
		\addlinespace
		IGR   &  40 & $4.1754\times10^{-4}$  & --   & $1.5208\times10^{-3}$  & --   & $6.3809\times10^{-4}$  & --   \\
		IGR   &  80 & $1.7243\times10^{-4}$  & 1.28 & $4.6808\times10^{-4}$  & 1.70 & $2.5167\times10^{-4}$  & 1.34 \\
		IGR   & 160 & $5.3812\times10^{-5}$  & 1.68 & $1.2836\times10^{-4}$  & 1.87 & $7.8125\times10^{-5}$  & 1.69 \\
		IGR   & 320 & $1.9085\times10^{-5}$  & 1.50 & $1.9629\times10^{-5}$  & 2.71 & $2.7988\times10^{-5}$  & 1.48 \\
		\bottomrule
	\end{tabular}
\end{table}

At $n=320$, the scalar-IGR density error is approximately $1.1\times 10^6$ times larger than the CoDeS error. CoDeS on the $n=80$ grid is already more accurate than scalar IGR on the $n=320$ grid in all the three primitive variables. The improvement is therefore not merely asymptotic but also gives a substantial fixed-resolution advantage. Because the exact solution remains smooth and expansive throughout the measurement interval, the degradation of scalar IGR cannot be attributed to shocks, contacts, or wave breaking. It is caused by stress activation in a region where the shock regularization should be inactive.

\subsection{Double Rarefaction}
\begin{figure}[t]
	\centering
	\begin{subfigure}[t]{0.48\textwidth}
		\centering
		\includegraphics[width=\linewidth]{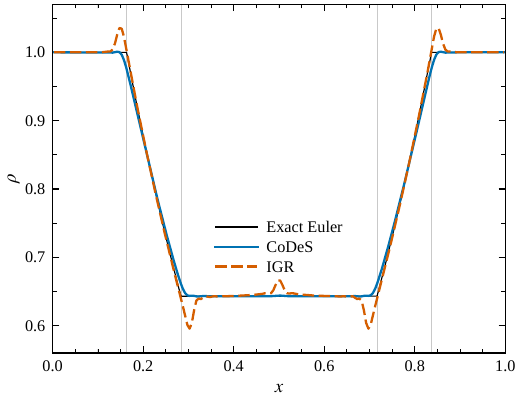}
		\caption{Density $\rho$.}
		\label{fig:double-rarefaction-rho}
	\end{subfigure}
	\hfill
	\begin{subfigure}[t]{0.48\textwidth}
		\centering
		\includegraphics[width=\linewidth]{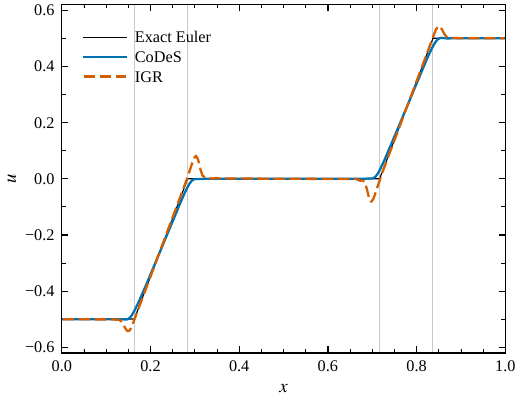}
		\caption{Velocity $u$.}
		\label{fig:double-rarefaction-u}
	\end{subfigure}
	
	\vspace{0.5em}
	
	\begin{subfigure}[t]{0.48\textwidth}
		\centering
		\includegraphics[width=\linewidth]{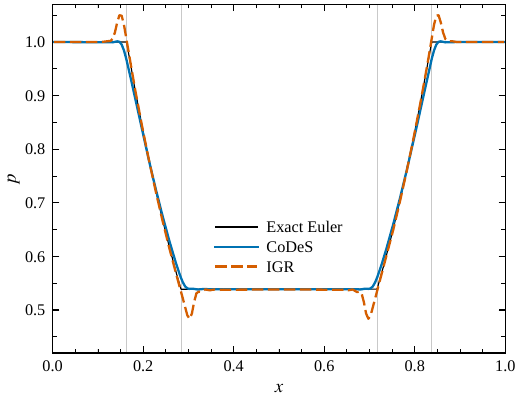}
		\caption{Pressure $p$.}
		\label{fig:double-rarefaction-p}
	\end{subfigure}
	\hfill
	\begin{subfigure}[t]{0.48\textwidth}
		\centering
		\includegraphics[width=\linewidth]{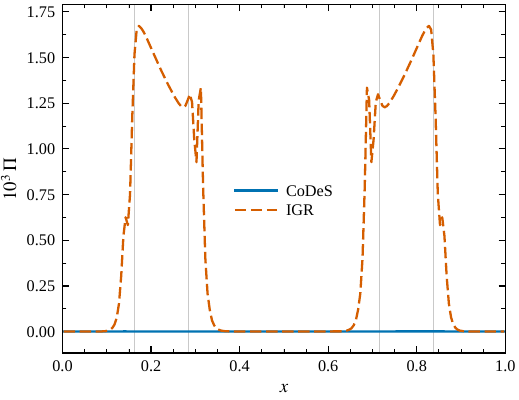}
		\caption{Entropic stress $\pi$.}
		\label{fig:double-rarefaction-pi}
	\end{subfigure}
	
	\caption{
		Double-rarefaction problem computed on the $n=200$ grid with seventh-order
		reconstruction and $C_{\alpha}=2$. The CoDeS and scalar-IGR solutions are
		compared with the exact Euler solution for $\rho$, $u$, and $p$, and the
		corresponding entropic stress is shown in panel~(d). CoDeS preserves the
		rarefaction structure without generating appreciable stress in the expansion
		regions, whereas scalar IGR produces nonzero stress near the rarefaction fans
		and correspondingly larger deviations in the primitive variables.
	}
	\label{fig:double-rarefaction-profiles}
\end{figure}

The second test is a double-rarefaction Riemann problem on $x\in[0,1]$. The initial discontinuity is located at $x_0=0.5$, with primitive states
\begin{equation}
	(\rho,u,p)_L=(1,-0.5,1),\qquad
	(\rho,u,p)_R=(1,0.5,1).
\end{equation}
The initial discontinuity is smoothed with width $\delta=h$. The calculation uses $n=200$ cells, final time $t=0.2$, extrapolation boundary conditions, and third-order strong-stability-preserving Runge--Kutta (SSP-RK3) time integration with
\begin{equation}
	\Delta t=\mathrm{CFL}\frac{h}{\max(|u|+a)},\qquad \mathrm{CFL}=0.15 .
\end{equation}
The exact solution consists of two rarefaction fans separated by a low-density intermediate state. Because the solution is non-smooth at the fan edges, this problem is not used as a formal convergence benchmark. Instead, it probes whether the regularization remains inactive within strong expansive gradients while preserving the Riemann structure.

Figure~\ref{fig:double-rarefaction-profiles} shows the density, velocity, pressure, and entropic stress at $t=0.2$. The CoDeS solution closely follows the exact rarefaction structure. The two fans are captured without visible overshoot, and the intermediate state is preserved in $u$, $p$, and $\rho$. This behavior is consistent with the one-dimensional compression gate. Both waves are expansive, so CoDeS supplies essentially no stress and the calculation reduces to the underlying high-order Euler discretization in the fan interiors.

Scalar IGR produces a finite entropic pressure across both rarefaction fans, with pronounced peaks near the fan edges. These stress peaks correlate with the overshoots and undershoots observed in the primitive variables, particularly near transitions between constant states and rarefaction fans. The contrast demonstrates a failure mode of deformation-magnitude-based regularization. Large gradients in rarefaction are not equivalent to shock-forming compression, and CoDeS avoids this false activation by conditioning the stress on compression rather than on gradient strength alone.

\subsection{Sod Shock Tube Case}

\begin{figure}[t]
	\centering
	\begin{subfigure}[t]{0.48\textwidth}
		\centering
		\includegraphics[width=\linewidth]{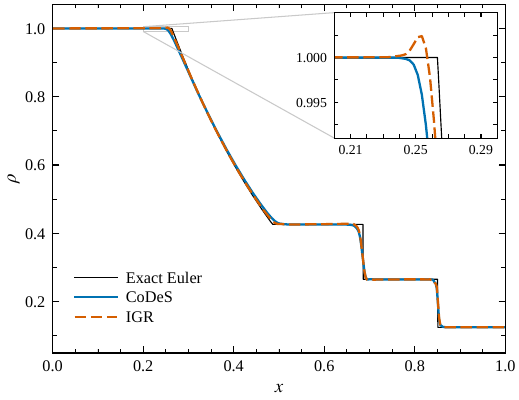}
		\caption{Density $\rho$.}
		\label{fig:sod-rho}
	\end{subfigure}
	\hfill
	\begin{subfigure}[t]{0.48\textwidth}
		\centering
		\includegraphics[width=\linewidth]{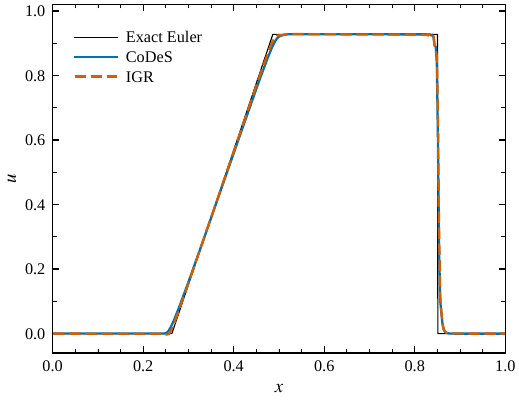}
		\caption{Velocity $u$.}
		\label{fig:sod-u}
	\end{subfigure}
	
	\vspace{0.5em}
	
	\begin{subfigure}[t]{0.48\textwidth}
		\centering
		\includegraphics[width=\linewidth]{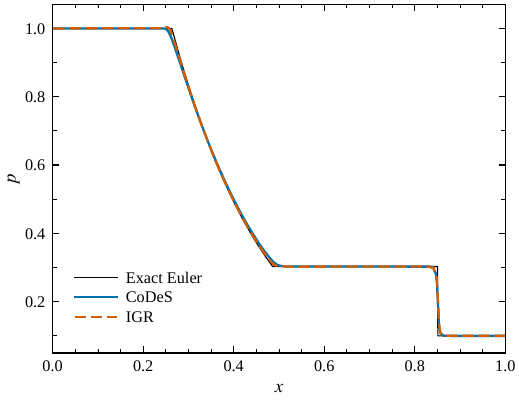}
		\caption{Pressure $p$.}
		\label{fig:sod-p}
	\end{subfigure}
	\hfill
	\begin{subfigure}[t]{0.48\textwidth}
		\centering
		\includegraphics[width=\linewidth]{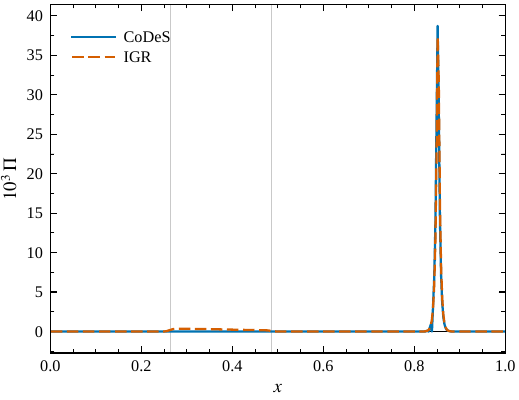}
		\caption{Entropic stress $\pi$.}
		\label{fig:sod-pi}
	\end{subfigure}
	
	\caption{
		Sod shock tube case at $t=0.2$ on the $n=400$ grid with $C_\alpha=2$,
		seventh-order linear upwind reconstruction, and SSP-RK3 time integration.
		The exact Euler solution is compared with CoDeS and scalar IGR for
		$\rho$, $u$, and $p$, and the corresponding entropic stress is shown in
		panel~(d). Both CoDeS and scalar IGR generate comparable stress at the
		right-going shock, confirming that the CoDeS compression gate does not
		suppress shock regularization. The difference appears in the non-shocking
		waves. CoDeS remains inactive in the left-going rarefaction and at the
		contact, whereas scalar IGR produces a weak but finite stress in the
		rarefaction region.
	}
	\label{fig:sod-profiles}
\end{figure}

The Sod shock tube case verifies that eliminating expansion activation does not remove the regularization needed at shocks. The problem is solved on $x\in[0,1]$ with $x_0=0.5$ and primitive states $(\rho,u,p)_L=(1,0,1)$ and $(\rho,u,p)_R=(0.125,0,0.1)$. The primitive variables are smoothed with width $\delta=2h$, and the solution is compared with the exact sharp Euler Riemann solution at $t=0.2$. The calculation uses $n=400$ cells, $C_\alpha=2$, extrapolation boundary conditions, seventh-order linear upwind reconstruction, and SSP-RK3 time integration with $\mathrm{CFL}=0.15$.

The exact solution contains a left-going rarefaction, a contact discontinuity, and a right-going shock. This makes the test a compact selectivity diagnostic. A successful closure should be inactive in the rarefaction and at the contact, but active at the shock. Figure~\ref{fig:sod-profiles} shows that both CoDeS and scalar IGR capture the main wave pattern and locate the rarefaction, contact, and shock correctly. Near the right-going shock the two methods generate comparable stress levels, confirming that the CoDeS compression gate does not suppress shock regularization. In the one-dimensional compressive region associated with the shock, CoDeS reduces to the compressive part of the IGR-type mechanism.

The difference appears in the non-shocking waves. Across the rarefaction, the velocity increases and the flow is locally expansive. CoDeS produces no visible entropic stress there, whereas scalar IGR produces a weak but finite stress over part of the fan. Although small compared with the shock stress, this nonzero response is visible in the density profile, where scalar IGR develops a slight overshoot near the rarefaction head. Similar overshoots appear at the same location in velocity and pressure. CoDeS also remains inactive at the contact, consistent with the nearly constant pressure and velocity across a density jump. The result therefore confirms the intended wave-type separation, with no stress in the rarefaction, no stress at the contact, and a localized stress peak at the shock.

\subsection{Two-dimensional Isentropic Vortex}

\begin{figure}[t]
	\centering
	\begin{subfigure}[t]{0.48\textwidth}
		\centering
		\includegraphics[width=\linewidth]{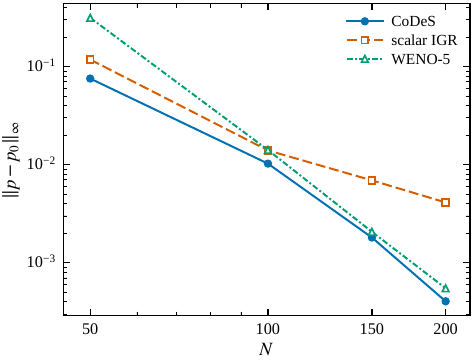}
		\caption{Pressure-error convergence after four periods.}
		\label{fig:isentropic-vortex-convergence}
	\end{subfigure}
	\hfill
	\begin{subfigure}[t]{0.48\textwidth}
		\centering
		\includegraphics[width=\linewidth]{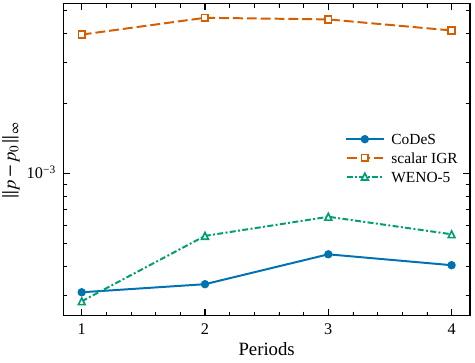}
		\caption{Pressure-error history on the $N=200$ mesh.}
		\label{fig:isentropic-vortex-error-growth}
	\end{subfigure}
	
	\caption{
		Pressure-error behavior for the two-dimensional isentropic vortex.
		Panel~(a) shows the $L^\infty$ pressure-error convergence after four
		periods, $t=400$. Panel~(b) shows the corresponding pressure-error history
		on the $N=200$ mesh over the first four periods. CoDeS maintains the lowest
		error over the tested resolutions and throughout the long-time advection,
		whereas scalar IGR exhibits substantially larger error growth in this smooth
		vortical flow.
	}
	\label{fig:isentropic-vortex-pressure-error}
\end{figure}

We next consider the standard two-dimensional isentropic vortex, a smooth periodic benchmark for numerical dissipation and long-time error growth. The flow contains no discontinuities and no shock-forming compression, so it isolates whether the regularization remains inactive in a smooth vortical field rather than acting as a generic artificial viscosity.

The vortex is initialized on the periodic domain $[-5,5]^2$ with free-stream state
\begin{equation}
	(\rho_\infty,p_\infty,u_\infty,v_\infty)=(1,1,0.1,0).
\end{equation}
Let $(x_c,y_c)=(0,0)$ denote the vortex center, $\tilde{x}=x-x_c$, $\tilde{y}=y-y_c$, and $r^2=\tilde{x}^2+\tilde{y}^2$. The isentropic vortex perturbation is prescribed as
\begin{equation}
	\delta u
	= -\frac{S}{2\pi}\tilde{y}
	\exp\!\left[\frac{\beta}{2}(1-r^2)\right],
	\qquad
	\delta v
	= \frac{S}{2\pi}\tilde{x}
	\exp\!\left[\frac{\beta}{2}(1-r^2)\right],
\end{equation}
and
\begin{equation}
	T
	= 1-\frac{(\gamma-1)S^2}{8\gamma\pi^2}
	\exp\!\left[\beta(1-r^2)\right],
	\qquad
	\rho=T^{1/(\gamma-1)},\qquad
	p=T^{\gamma/(\gamma-1)} .
\end{equation}
The primitive velocity is then
\begin{equation}
	u=u_\infty+\delta u,\qquad v=v_\infty+\delta v .
\end{equation}
Here $S$ is the vortex-strength parameter. It sets the amplitude of the swirling velocity perturbation through the isentropic relation. This velocity perturbation then determines the corresponding density and pressure deficit in the vortex core. The parameter $\beta$ controls the radial decay of the vortex profile. In this test we use $S=5$ and $\beta=1$. The exact solution is a rigid translation at velocity $(u_\infty,v_\infty)$, and one period is $T=10/u_\infty=100$. Errors are measured after four periods, at $t=400$, by comparing the numerical solution with the initial condition shifted by the exact translation. Conservative initial cell averages are evaluated with a tensor-product three-point Gauss-Legendre quadrature.

Uniform meshes with $N^2=50^2,100^2,150^2,200^2$ cells are employed. All methods use periodic boundary conditions, the same LF/Rusanov numerical flux, and the same final time. CoDeS and scalar IGR use fifth-order linear finite-volume reconstruction with $\alpha=2h^2$, while the comparison calculation uses WENO-5 reconstruction with the LF/Rusanov flux, as implemented in the open-source Multiphase Flow Solver MFC \cite{bryngelson2021mfc}.

Table~\ref{tab:isentropic-vortex-pressure-linf} reports the pressure $L^\infty$ error after four periods. CoDeS gives the smallest error at every tested resolution. At $N=200$, the CoDeS error is $4.0431\times10^{-4}$, compared with $5.4890\times10^{-4}$ for WENO-5/LF and $4.1168\times10^{-3}$ for scalar IGR. On the finest mesh, CoDeS is therefore approximately $1.36$ times more accurate than WENO-5/LF and $10.2$ times more accurate than scalar IGR.

\begin{table}[h]
	\centering
	\small
	\caption{
		Pressure $L^\infty$ errors for the two-dimensional isentropic vortex after
		four periods, $t=400$. The observed rates are computed between successive
		mesh resolutions.
	}
	\label{tab:isentropic-vortex-pressure-linf}
	\setlength{\tabcolsep}{4pt}
	\begin{tabular}{@{}l c c c c c c c@{}}
		\toprule
		Method & $N=50$ & Rate & $N=100$ & Rate & $N=150$ & Rate & $N=200$ \\
		\midrule
		CoDeS & $7.5511\times10^{-2}$ & 2.88 & $1.0251\times10^{-2}$ & 4.28 &  $1.8079\times10^{-3}$ & 5.21 & $4.0431\times10^{-4}$ \\
		scalar IGR & $1.1794\times10^{-1}$ & 3.08 & $1.3983\times10^{-2}$ & 1.74 & $6.9157\times10^{-3}$ & 1.80 & $4.1168\times10^{-3}$ \\
		WENO-5/LF & $3.1455\times10^{-1}$ & 4.48 & $1.4114\times10^{-2}$ & 4.74 & $2.0679\times10^{-3}$ & 4.61 & $5.4890\times10^{-4}$ \\
		\bottomrule
	\end{tabular}
\end{table}

Figure~\ref{fig:isentropic-vortex-pressure-error} compares the behavior of the three methods on this case. Figure~\ref{fig:isentropic-vortex-convergence} shows the pressure-error convergence. CoDeS follows the expected high-order trend and remains slightly more accurate than WENO-5/LF over the tested range. Scalar IGR improves from $N=50$ to $N=100$, but its convergence deteriorates on the finer meshes, with rates below second order over the last two refinement intervals. Figure~\ref{fig:isentropic-vortex-error-growth} shows the pressure-error history at $N=200$ over the first four periods. CoDeS maintains the lowest error throughout the long-time advection. WENO-5/LF reports a close but consistently larger error after the first period, whereas scalar IGR deviates from both methods from the first period onward.

The computational overhead of CoDeS is modest in this benchmark. On the CPU platform used for these runs, CoDeS costs between $4\%$ and $20\%$ more wall time than the MFC WENO-5/LF reference over the tested grids, while scalar IGR costs between $0\%$ and $10\%$ more. On the finest mesh, CoDeS requires $679$ s, compared with $623$ s for scalar IGR and $564$ s for WENO-5/LF. Because the CoDeS/IGR and WENO-5/LF results are obtained through different code paths, these timings should be interpreted as practical wall-clock comparisons rather than strict algorithm-only performance measurements.

The error behavior follows from the activation mechanisms. The analytic vortex is smooth and essentially divergence-free, so a shock regularization should generate little stress. On the $N=50$ initial field, the CoDeS source is small,
\begin{equation}
	\|b_{\mathrm{CoDeS}}\|_{\infty}=2.72\times10^{-3},\qquad
	\|b_{\mathrm{CoDeS}}\|_{1}=1.26\times10^{-4},
\end{equation}
consistent with finite-grid differentiation and projection effects. By contrast, the scalar IGR source is strongly sign-indefinite,
\begin{equation}
	\min Q_{\mathrm{IGR}}=-3.17,\qquad
	\max Q_{\mathrm{IGR}}=4.47\times10^{-1}.
\end{equation}
This sign-indefiniteness reflects the rotational contribution in the multidimensional scalar IGR source and explains its false activation in smooth vortical flows. The vortex test therefore shows that CoDeS does not achieve robustness by indiscriminately adding dissipation. In smooth rotational flows, the tensor regularization remains weak and the method largely recovers the underlying high-order finite-volume discretization.

\subsection{Perturbed Two-dimensional Riemann Problem}

\begin{figure}[t]
	\centering
	\begin{subfigure}[t]{0.32\textwidth}
		\centering
		\includegraphics[width=\linewidth]{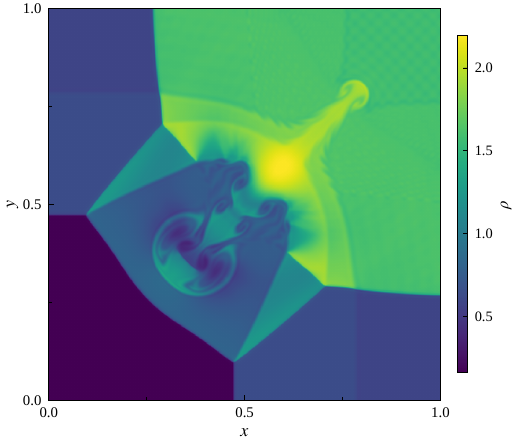}
		\caption{CoDeS.}
		\label{fig:riemann-density-codes}
	\end{subfigure}
	\hfill
	\begin{subfigure}[t]{0.32\textwidth}
		\centering
		\includegraphics[width=\linewidth]{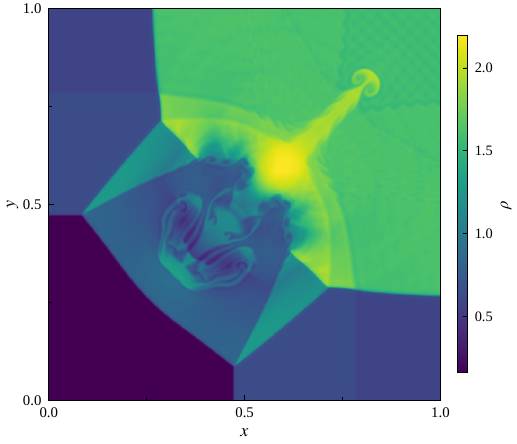}
		\caption{Scalar IGR.}
		\label{fig:riemann-density-igr}
	\end{subfigure}
	\hfill
	\begin{subfigure}[t]{0.32\textwidth}
		\centering
		\includegraphics[width=\linewidth]{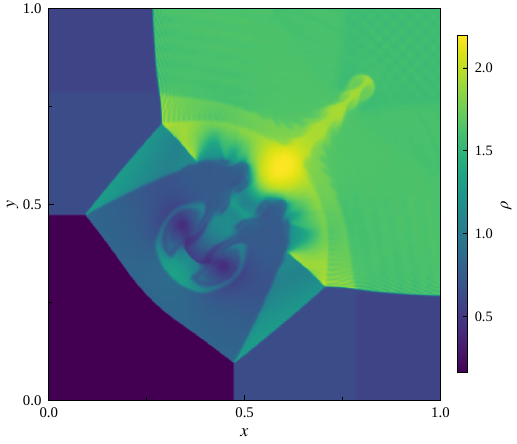}
		\caption{WENO-5/LF.}
		\label{fig:riemann-density-weno}
	\end{subfigure}
	
	\caption{
		Density fields for the perturbed two-dimensional Riemann problem at
		$t=0.8$ on a $500\times500$ grid. All three calculations reproduce the
		large-scale interacting Riemann structure. CoDeS preserves a sharper and
		more coherent central roll-up while maintaining clean shock transitions,
		whereas scalar IGR visibly distorts parts of the vortical interaction region
		and WENO-5/LF gives a more diffuse roll-up with oscillatory artifacts near
		some shock fronts.
	}
	\label{fig:riemann-density-comparison}
\end{figure}

\begin{figure}[t]
	\centering
	\begin{subfigure}[t]{0.48\textwidth}
		\centering
		\includegraphics[width=\linewidth]{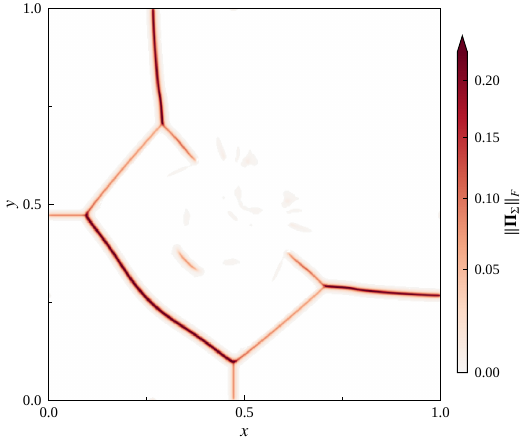}
		\caption{CoDeS stress magnitude $\|\boldsymbol{\Pi}_{\Sigma}\|_F$.}
		\label{fig:riemann-codes-stress}
	\end{subfigure}
	\hfill
	\begin{subfigure}[t]{0.48\textwidth}
		\centering
		\includegraphics[width=\linewidth]{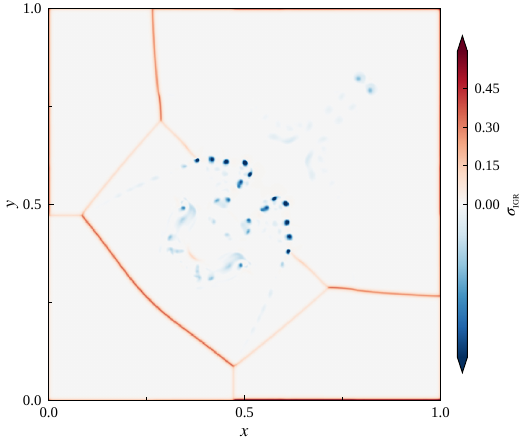}
		\caption{Scalar IGR entropic pressure $\sigma_{\mathrm{IGR}}$.}
		\label{fig:riemann-igr-pressure}
	\end{subfigure}
	
	\caption{
		Activation diagnostics for the perturbed two-dimensional Riemann problem at
		$t=0.8$. The CoDeS stress magnitude is concentrated primarily along shocks
		and strong compressive wave fronts, while remaining small in the central
		shear-layer roll-up and over most smooth regions. The scalar IGR entropic
		pressure activates more broadly, including in vortical and shear-dominated
		regions, and exhibits sign-indefinite patches inside the roll-up.
	}
	\label{fig:riemann-activation-diagnostics}
\end{figure}

The next test is a perturbed two-dimensional Riemann problem on $[0,1]\times[0,1]$. It contains interacting shocks, contact discontinuities, slip lines, and shear-layer roll-up in the central interaction region, which together provide a stringent multidimensional test of shock regularization, contact preservation, and vortical-structure resolution.

The inviscid Euler equations are solved with four constant states separated at $x_0=y_0=0.75$, as listed in Table~\ref{tab:riemann-2d-initial-states}.

\begin{table}[h!]
	\centering
	\small
	\caption{
		Initial quadrant states for the perturbed two-dimensional Riemann problem.
		The four constant states are separated at $x_0=y_0=0.75$ before the density
		perturbation is applied.
	}
	\label{tab:riemann-2d-initial-states}
	\setlength{\tabcolsep}{7pt}
	\begin{tabular}{@{}l c c c c@{}}
		\toprule
		Region & $\rho_0$ & $u$ & $v$ & $p$ \\
		\midrule
		$x<x_0,\ y<y_0$ & $0.138$  & $1.206$ & $1.206$ & $0.029$ \\
		$x<x_0,\ y>y_0$ & $0.5323$ & $1.206$ & $0$     & $0.3$   \\
		$x>x_0,\ y<y_0$ & $0.5323$ & $0$     & $1.206$ & $0.3$   \\
		$x>x_0,\ y>y_0$ & $1.5$    & $0$     & $0$     & $1.5$   \\
		\bottomrule
	\end{tabular}
\end{table}

To seed shear-layer instability and probe small-scale structure preservation, the density is perturbed as $\rho(x,y,0)=\rho_0(x,y)+0.025+0.025\sin(50\pi x)\sin(50\pi y)$, while velocity and pressure are unperturbed. For the CoDeS and scalar IGR calculations, the quadrant jumps are mildly smoothed in primitive variables before the density perturbation is added. The WENO-5/LF reference calculation uses the same four states and perturbation but keeps the quadrant interfaces sharp.

All computations use a $500\times500$ uniform grid with zero-gradient extrapolation boundaries. The final time is $t=0.8$, reached with $\Delta t=8\times10^{-5}$ over $10000$ time steps. CoDeS and scalar IGR use fifth-order finite-volume reconstruction, SSP-RK3 time integration, and $\alpha=2h^2$. The modified-Helmholtz solve uses five iterations per Runge--Kutta stage after an initial $150$-iteration warm start. The WENO reference uses fifth-order WENO reconstruction with the LF/Rusanov flux.

Figure~\ref{fig:riemann-density-comparison} compares the final-time density fields. All the three calculations reproduce the large-scale Riemann structure, including the interacting shocks, contact surfaces, and central shear-layer roll-up. The detailed morphology, however, differs. WENO-5/LF captures the principal wave pattern but yields a more diffuse central vortical region and visible oscillatory artifacts near several shock fronts. Scalar IGR suppresses some shock-associated oscillations, but it also distorts the central roll-up, particularly in the lower-left portion of the interaction region near $(x,y)\approx(0.3,0.4)$. CoDeS preserves a sharper and more coherent roll-up while maintaining clean shock transitions.

The activation diagnostics in Figure~\ref{fig:riemann-activation-diagnostics} explain this difference. The scalar IGR entropic pressure $\sigma_{\mathrm{IGR}}$ is activated not only along shocks and compressive wave fronts but also in the central vortical region, where negative patches appear inside the roll-up. These regions are dominated by shear and rotation rather than by shock-like compression, and their correlation with distorted density structures indicates that scalar IGR interacts unfavorably with the vortical dynamics. The CoDeS stress magnitude $\|\boldsymbol{\Pi}_{\Sigma}\|_F$, by contrast, concentrates along shocks and strong compressive wave fronts. It remains small in the central roll-up, near contacts, and over most of the smooth regions. CoDeS therefore changes not only the amount of regularization but also its location, localizing the stress to genuinely compressive structures rather than to arbitrary multidimensional deformation.

This test demonstrates the multidimensional selectivity of CoDeS. Like scalar IGR, it supplies regularization at shocks and reduces shock-induced artifacts. Unlike scalar IGR, it avoids spurious stress in the shear- and vorticity-dominated roll-up. Compared with WENO-5/LF, it preserves sharper vortical structure while maintaining stable shock transitions, providing a more favorable balance between shock regularization and fine-scale flow resolution.

\subsection{Viscous Shock-tube Problem}

\begin{figure}[t]
	\centering
	\begin{subfigure}[t]{0.48\textwidth}
		\centering
		\includegraphics[width=\linewidth]{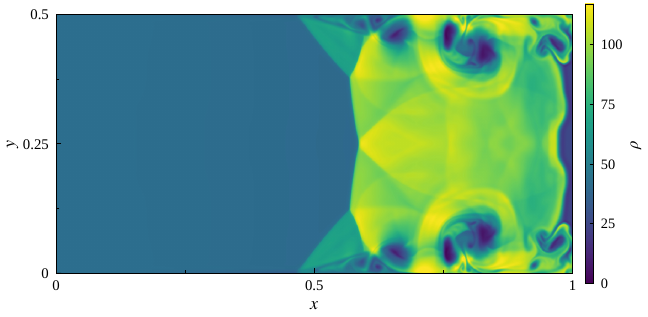}
		\caption{Density $\rho$.}
		\label{fig:vst-density}
	\end{subfigure}
	\hfill
	\begin{subfigure}[t]{0.48\textwidth}
		\centering
		\includegraphics[width=\linewidth]{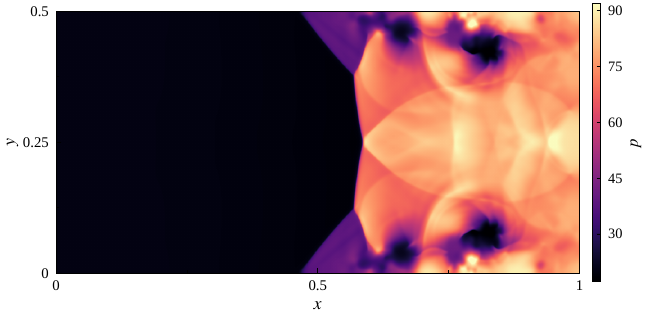}
		\caption{Pressure $p$.}
		\label{fig:vst-pressure}
	\end{subfigure}
	
	\vspace{0.5em}
	
	\begin{subfigure}[t]{0.48\textwidth}
		\centering
		\includegraphics[width=\linewidth]{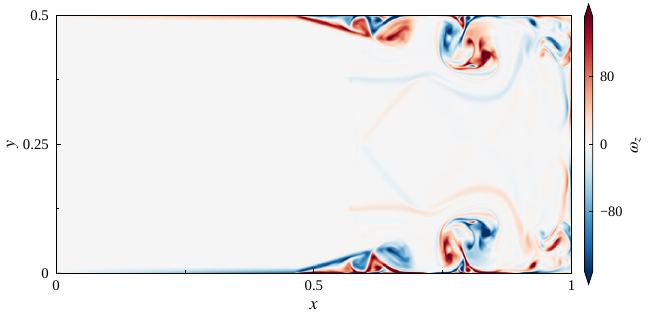}
		\caption{Spanwise vorticity $\omega_z$.}
		\label{fig:vst-vorticity}
	\end{subfigure}
	\hfill
	\begin{subfigure}[t]{0.48\textwidth}
		\centering
		\includegraphics[width=\linewidth]{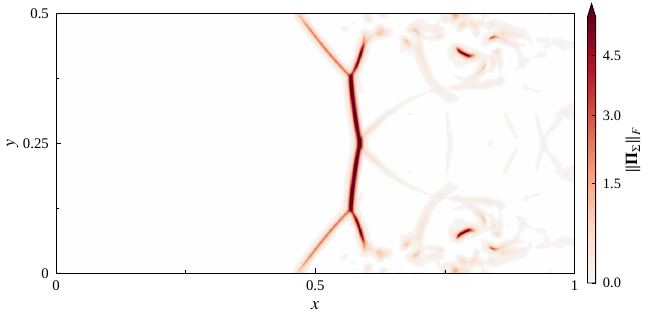}
		\caption{CoDeS stress magnitude $\|\boldsymbol{\Pi}_{\Sigma}\|_F$.}
		\label{fig:vst-codes-stress}
	\end{subfigure}
	
	\caption{
		Viscous shock-tube problem at $t=1.0$ on the $1280\times640$ grid.
		Shown are density $\rho$, pressure $p$, spanwise vorticity $\omega_z$,
		and the CoDeS stress magnitude $\|\boldsymbol{\Pi}_{\Sigma}\|_F$.
		The density and pressure fields show the reflected shock system, oblique
		compression waves, and post-shock structures generated by shock-wall and
		shock--boundary-layer interaction. The vorticity field highlights the
		wall-generated shear layers and roll-up structures, while the CoDeS stress
		remains concentrated primarily along the main compression front and attached
		oblique branches.
	}
	\label{fig:vst-fields}
\end{figure}

We next consider a two-dimensional viscous shock-tube problem. This case combines shock propagation and reflection, wall-generated boundary layers, shear-layer roll-up, and shock--vortex interaction. It therefore tests whether CoDeS remains localized to compressive wave structures while preserving the vortical dynamics generated by physical viscosity and no-slip walls.

The compressible Navier--Stokes equations are solved on $(x,y)\in[0,1]\times[0,0.5]$ with no-slip wall boundary conditions on all four sides. The mesh contains $1280\times640$ uniform cells, and the dynamic viscosity is $\mu=10^{-3}$. Time integration uses the third-order strong-stability-preserving Runge--Kutta method (SSP-RK3) with adaptive time stepping. At each step, $\Delta t$ is chosen to satisfy the prescribed target CFL number, $\mathrm{CFL}=0.1$ \cite{TVDRK}, including the relevant convective and viscous stability constraints. The final time is $t=1.0$.

The initial condition consists of two quiescent states separated by a vertical diaphragm at $x_0=0.5$,
\begin{equation}
	(\rho,u,v,p)(x,0)=
	\begin{cases}
		\left(120,0,0,\dfrac{120}{\gamma}\right), & x < x_0,\\[4pt]
		\left(1.2,0,0,\dfrac{1.2}{\gamma}\right), & x > x_0.
	\end{cases}
\end{equation}
The diaphragm is smoothed in the numerical initial condition. The CoDeS calculation uses $\boldsymbol{\Pi}_{\Sigma}=\sigma\boldsymbol{M}$ with $\alpha=10h^2$. A third-order finite-volume reconstruction is employed, and the modified-Helmholtz equation for $\sigma$ is solved with five iterations per Runge--Kutta stage after an initial warm start. No auxiliary shock-localized face source is used, so the entropic tensor is determined entirely by the baseline compression-directional source. The tensor stress contribution is projected consistently at solid boundaries.

Figure~\ref{fig:vst-fields} shows density, pressure, spanwise vorticity $\omega_z$, and $\|\boldsymbol{\Pi}_{\Sigma}\|_F$ at $t=1.0$. The density field displays the expected complex shock-tube structure following shock--boundary-layer interaction, including reflected compression systems, oblique waves, high-density post-shock regions, and pronounced vortical roll-up near the upper and lower walls. The wall-induced vortices remain sharply visible, indicating that the entropic stress does not simply smear shear layers or suppress no-slip-boundary-layer instability.

The pressure field provides a complementary view of the compressive wave pattern. It is smoother than the density field, as expected in a viscous compressible calculation, whereas it clearly identifies the main shock and post-shock compression regions. No prominent grid-scale pressure oscillations are visible near the principal compression fronts. The vorticity field, by contrast, isolates the shear-dominated dynamics, with large positive and negative values of $\omega_z$ concentrated near the walls and in the roll-up structures generated behind the shock.

The CoDeS stress distribution differs markedly from the vorticity distribution. The strongest $\|\boldsymbol{\Pi}_{\Sigma}\|_F$ response concentrates along the main compression front and its attached oblique branches, while the stress remains weak in most wall-generated vortical roll-up regions. Some localized stress also appears in the post-shock flow, but these patches are confined to the regions adjacent to sharp pressure gradients. They indicate the presence of secondary shock-wave structures in those regions, where CoDeS is activated appropriately to regularize the local compressive dynamics.

The viscous shock tube demonstrates that CoDeS can operate in a regime where shocks and intense vortical structures coexist. The method supplies regularization at shock and compression fronts, where under-resolved compressive dynamics require stabilization, while leaving wall-generated vorticity and shear-layer roll-up largely unaffected. This separation is essential in viscous compressible flows, since excessive activation in vortical regions would compromise the physically relevant post-shock dynamics.

\subsection{Two-fluid Triple-point Problem}
\begin{figure}[t]
	\centering
	\begin{subfigure}[t]{0.48\textwidth}
		\centering
		\includegraphics[width=\linewidth]{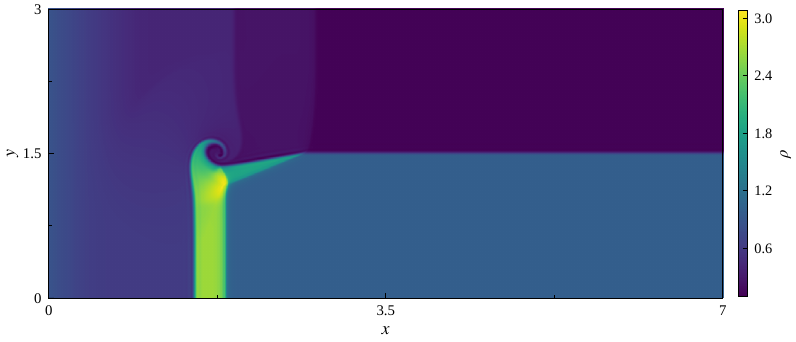}
		\caption{$t=1$.}
		\label{fig:tp-density-T1}
	\end{subfigure}
	\hfill
	\begin{subfigure}[t]{0.48\textwidth}
		\centering
		\includegraphics[width=\linewidth]{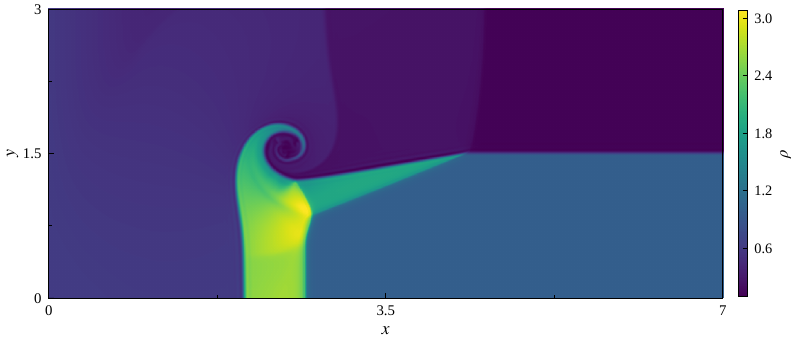}
		\caption{$t=2$.}
		\label{fig:tp-density-T2}
	\end{subfigure}
	
	\vspace{0.5em}
	
	\begin{subfigure}[t]{0.48\textwidth}
		\centering
		\includegraphics[width=\linewidth]{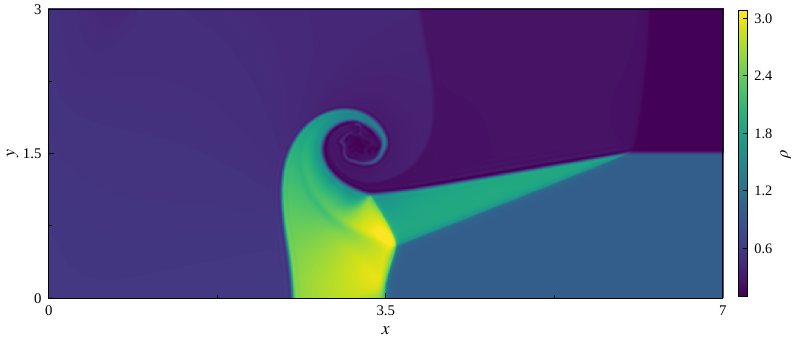}
		\caption{$t=3$.}
		\label{fig:tp-density-T3}
	\end{subfigure}
	\hfill
	\begin{subfigure}[t]{0.48\textwidth}
		\centering
		\includegraphics[width=\linewidth]{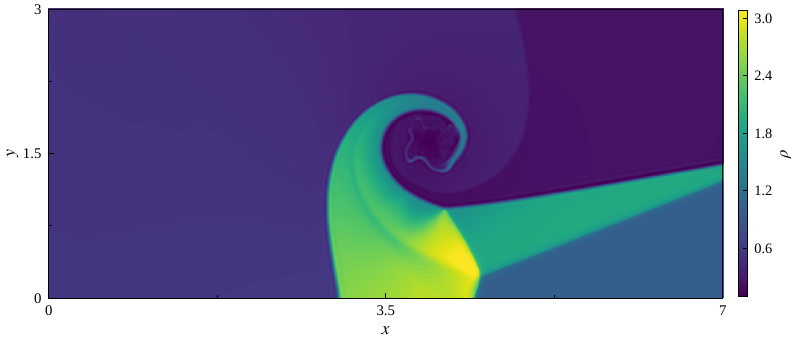}
		\caption{$t=4$.}
		\label{fig:tp-density-T4}
	\end{subfigure}
	
	\caption{
		Density evolution for the two-fluid triple-point problem at
		$t=1,2,3$, and $4$. The sequence shows the deformation of the material
		interface, the development of baroclinically generated roll-up, and the
		propagation of the associated compressive wave system.
	}
	\label{fig:tp-density-evolution}
\end{figure}

\begin{figure}[t]
	\centering
	\begin{subfigure}[t]{0.48\textwidth}
		\centering
		\includegraphics[width=\linewidth]{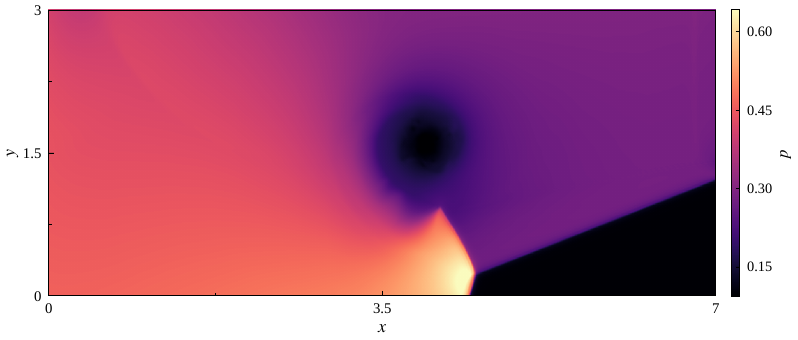}
		\caption{Pressure $p$.}
		\label{fig:tp-pressure-T4}
	\end{subfigure}
	\hfill
	\begin{subfigure}[t]{0.48\textwidth}
		\centering
		\includegraphics[width=\linewidth]{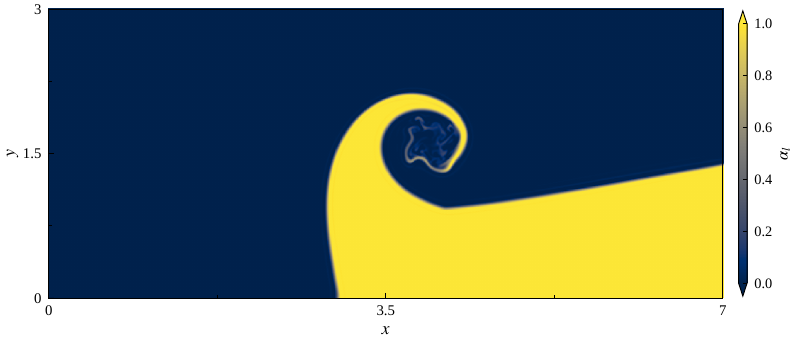}
		\caption{Volume fraction $\alpha_\ell$.}
		\label{fig:tp-alpha-T4}
	\end{subfigure}
	
	\vspace{0.5em}
	
	\begin{subfigure}[t]{0.48\textwidth}
		\centering
		\includegraphics[width=\linewidth]{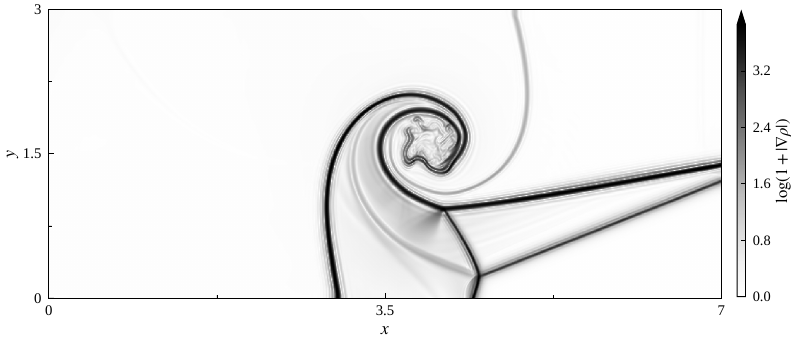}
		\caption{Density-gradient indicator $\log(1+|\nabla\rho|)$.}
		\label{fig:tp-density-gradient-T4}
	\end{subfigure}
	\hfill
	\begin{subfigure}[t]{0.48\textwidth}
		\centering
		\includegraphics[width=\linewidth]{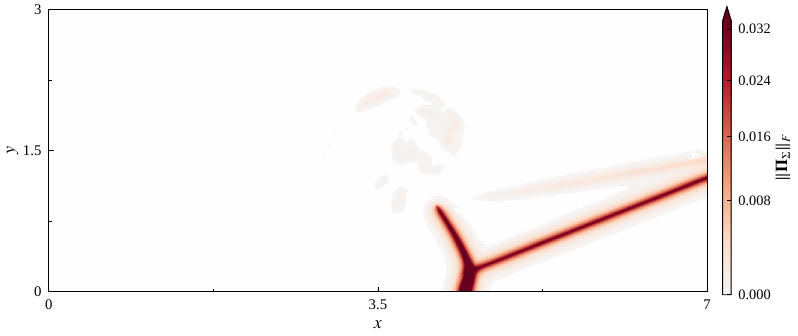}
		\caption{CoDeS stress magnitude $\|\boldsymbol{\Pi}_{\Sigma}\|_F$.}
		\label{fig:tp-codes-stress-T4}
	\end{subfigure}
	
	\caption{
		Final-time diagnostics for the two-fluid triple-point problem at $t=4$.
		Shown are pressure $p$, volume fraction $\alpha_\ell$, the density-gradient
		indicator $\log(1+|\nabla\rho|)$, and the CoDeS stress magnitude
		$\|\boldsymbol{\Pi}_{\Sigma}\|_F$. The pressure and density-gradient fields
		identify the compressive waves and material interfaces, while the stress
		diagnostic shows that CoDeS is concentrated primarily along compressive
		wave structures and remains weak over most of the rolled-up material
		interface.
	}
	\label{fig:tp-final-diagnostics}
\end{figure}

The two-fluid triple-point problem is a demanding multi-material test because it couples shock propagation, material-interface deformation, baroclinic vorticity generation, inviscid shear-layer roll-up, and repeated wave--interface interaction. It probes whether CoDeS can stabilize compressive structures without diffusing material interfaces or suppressing interface-driven vortical dynamics.

The inviscid two-fluid compressible Euler equations are solved on $(x,y)\in[0,7]\times[0,3]$ using a $700\times300$ uniform grid and ghost-cell extrapolation boundaries. Time integration is performed with SSP-RK3, adaptive time stepping, and target $\mathrm{CFL}=0.8$. The final time is $t=4.0$, and solution snapshots are saved every $\Delta t=0.04$. The two ideal gases have $\gamma_1=1.5$ and $\gamma_2=1.4$, consequently $\Gamma_1=1/(\gamma_1-1)=2.0$ and $\Gamma_2=1/(\gamma_2-1)=2.5$.

The initial data consist of a high-pressure left state and two low-pressure right states,
\begin{equation}
	\begin{aligned}
		(\rho,u,v,p,\alpha_\ell)_L  &= (1,     0,0,1,   1),\\
		(\rho,u,v,p,\alpha_\ell)_T  &= (0.125, 0,0,0.1, 1),\\
		(\rho,u,v,p,\alpha_\ell)_B  &= (1,     0,0,0.1, 0).
	\end{aligned}
\end{equation}
Here $\alpha_\ell\equiv\alpha_1$ denotes the volume fraction of fluid 1, so $\alpha_\ell=1$ corresponds to a state dominated by fluid 1 and $\alpha_\ell=0$ to a state dominated by fluid 2. The left and upper-right states therefore contain fluid 1, while the lower-right state contains fluid 2.

In the numerical initialization, volume-fraction guard values of order $10^{-8}$ are used to avoid exactly vanishing phase fractions. The plotted mixture density is
\begin{equation}
	\rho=\alpha_1\rho_1+\alpha_2\rho_2,\qquad \alpha_2=1-\alpha_1,
\end{equation}
with the smoothed quadrant weights applied to the conservative phase-density and volume-fraction variables.

The CoDeS calculation uses the baseline tensor entropic-stress closure with $\alpha=10h^2$, where $h=\max(\Delta x,\Delta y)$. Both the hyperbolic flux and the tensor regularization construction use seventh-order reconstruction. The modified-Helmholtz equation for $\sigma$ is solved with three iterations per Runge--Kutta stage after the initial warm start. No auxiliary face-normal source or boundary fallback is used, so the regularization shown here is the baseline CoDeS tensor stress.

Figure~\ref{fig:tp-density-evolution} shows the density field at $t=1,2,3$, and $4$. The sequence illustrates the development of the triple-point interaction from the initial pressure and material interfaces. The high-pressure left state drives a wave system into the two right states, while the material interface begins to deform near the triple point. Baroclinic vorticity deposition and inviscid velocity shear then generate a pronounced roll-up. By $t=4$, the interface has formed a coherent spiral with embedded fine-scale structures, while the outgoing compression fronts and material boundaries remain sharply defined.

This density evolution is significant because the problem contains both strong compressive waves and inviscid interface dynamics. In the absence of physical viscosity, the small-scale roll-up is controlled by the Euler dynamics, the initial smoothing, the mesh, and the numerical regularization. An isotropic artificial viscosity would tend to smear the spiral and damp thin interfacial structures. The CoDeS result, by contrast, retains a clearly visible material-interface roll-up, preserves thin features in the central interaction region, and maintains a clean large-scale wave pattern without dominant grid-scale pressure noise.

Figure~\ref{fig:tp-final-diagnostics} presents the final-time pressure, volume fraction $\alpha_\ell$, density-gradient indicator $\log(1+|\nabla\rho|)$, and $\|\boldsymbol{\Pi}_{\Sigma}\|_F$. The pressure field shows no strong post-shock noise, indicating adequate regularization of compressible dynamics. The volume-fraction field remains nearly binary in most of the domain, and the material interface remains sharply defined after substantial deformation. The density-gradient indicator gives a Schlieren-type view in which the largest values occur along material interfaces, shock fronts, and thin roll-up filaments.

The entropic-stress diagnostic provides the main evidence of selectivity. The field $\|\boldsymbol{\Pi}_{\Sigma}\|_F$ is concentrated primarily along the outgoing compressive waves and the localized compression regions generated by the triple-point interaction. It remains weak over most of the rolled-up material interface and in the smooth regions surrounding the central vortex. CoDeS is therefore not activated merely because the solution contains a large density gradient, a sharp volume-fraction transition, or inviscid vortical roll-up. It is activated primarily where the velocity field contains compressive strain.

Overall, the two-fluid triple-point calculation demonstrates robustness in a genuinely multidimensional multi-material flow. The result shows stable shock--interface interaction, well-preserved inviscid roll-up, sharp material interfaces, and localized compressive regularization. This is the balance required in multispecies compressible applications such as supersonic combustion, where shocks must be regularized without excessive diffusion of composition gradients, contact surfaces, or vortical mixing structures. Direct assessment in reacting multispecies flows is left for future work.

\subsection{Mach-3 Slot Jet with Boundary-localized Flux Fallback}
\begin{figure}[t]
	\centering
	\begin{subfigure}[t]{0.48\textwidth}
		\centering
		\includegraphics[width=\linewidth]{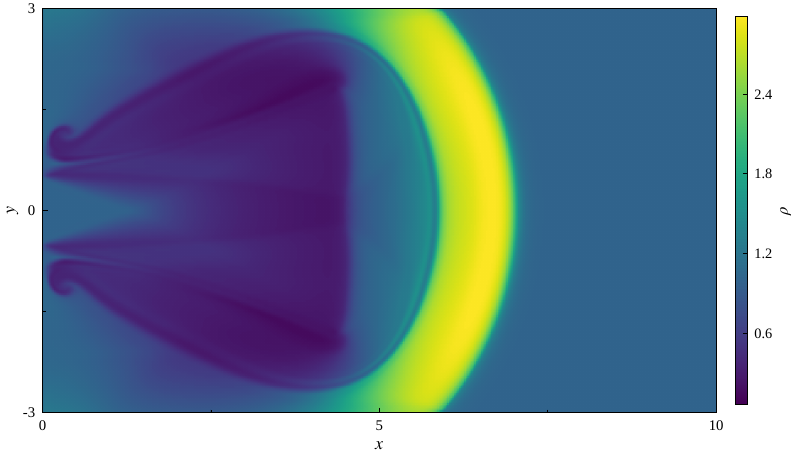}
		\caption{$t=2$.}
		\label{fig:slot-jet-density-T2}
	\end{subfigure}
	\hfill
	\begin{subfigure}[t]{0.48\textwidth}
		\centering
		\includegraphics[width=\linewidth]{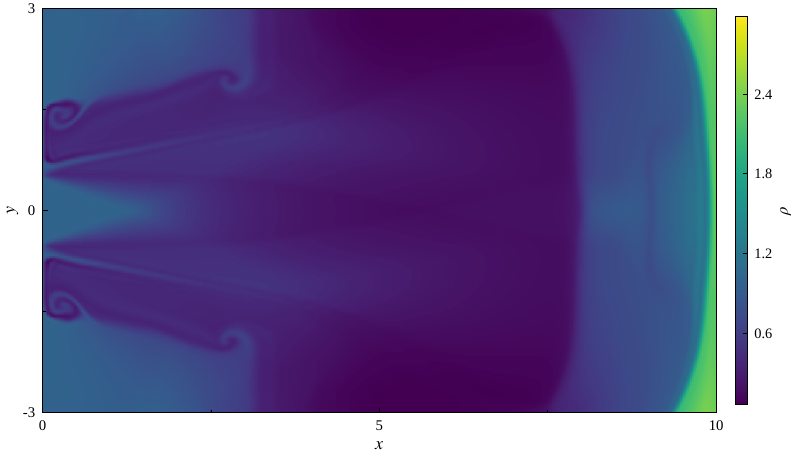}
		\caption{$t=4$.}
		\label{fig:slot-jet-density-T4}
	\end{subfigure}
	
	\vspace{0.5em}
	
	\begin{subfigure}[t]{0.48\textwidth}
		\centering
		\includegraphics[width=\linewidth]{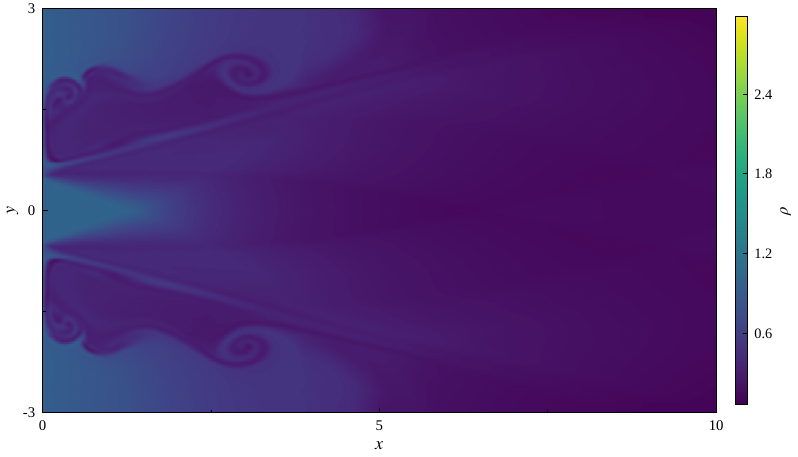}
		\caption{$t=6$.}
		\label{fig:slot-jet-density-T6}
	\end{subfigure}
	\hfill
	\begin{subfigure}[t]{0.48\textwidth}
		\centering
		\includegraphics[width=\linewidth]{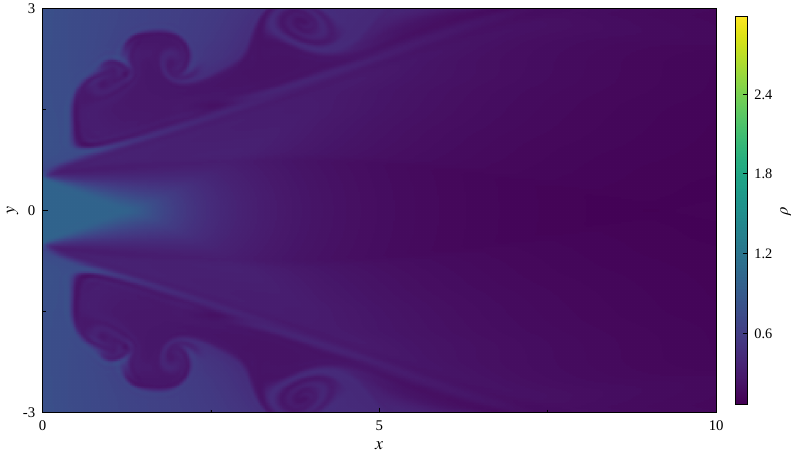}
		\caption{$t=8$.}
		\label{fig:slot-jet-density-T8}
	\end{subfigure}
	
	\caption{
		Density evolution for the Mach--3 slot jet at $t=2,4,6$, and $8$.
		The sequence shows the formation of the jet core, shock-cell structure,
		barrel-shock system, and shear-layer roll-up while the plume remains stable
		through the time duration.
	}
	\label{fig:slot-jet-density-evolution}
\end{figure}

\begin{figure}[h]
	\centering
	\begin{subfigure}[t]{0.48\textwidth}
		\centering
		\includegraphics[width=\linewidth]{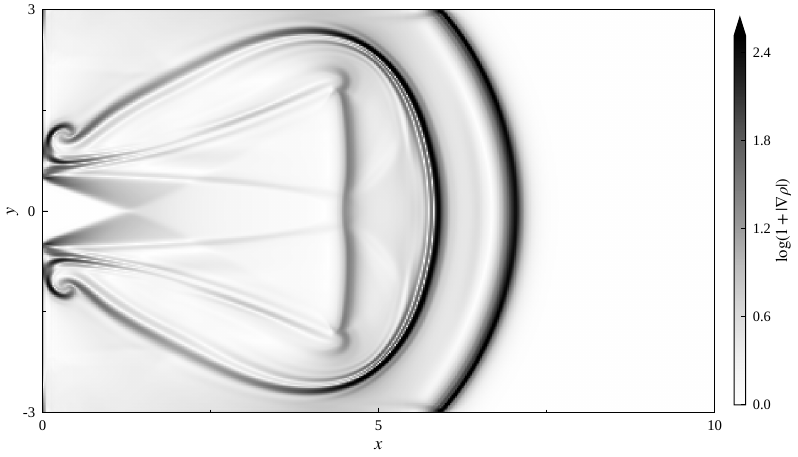}
		\caption{$t=2$.}
		\label{fig:slot-jet-density-gradient-T2}
	\end{subfigure}
	\hfill
	\begin{subfigure}[t]{0.48\textwidth}
		\centering
		\includegraphics[width=\linewidth]{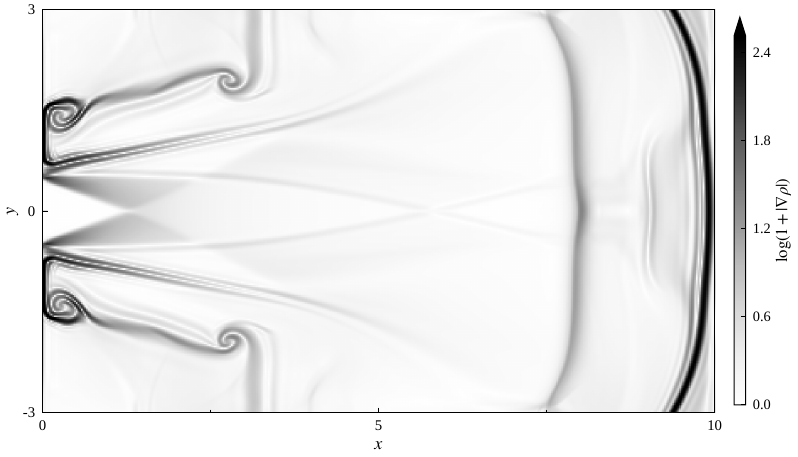}
		\caption{$t=4$.}
		\label{fig:slot-jet-density-gradient-T4}
	\end{subfigure}
	
	\vspace{0.5em}
	
	\begin{subfigure}[t]{0.48\textwidth}
		\centering
		\includegraphics[width=\linewidth]{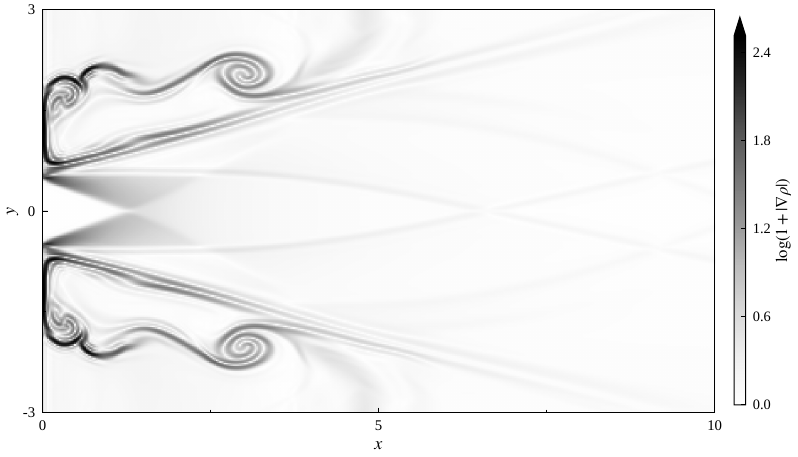}
		\caption{$t=6$.}
		\label{fig:slot-jet-density-gradient-T6}
	\end{subfigure}
	\hfill
	\begin{subfigure}[t]{0.48\textwidth}
		\centering
		\includegraphics[width=\linewidth]{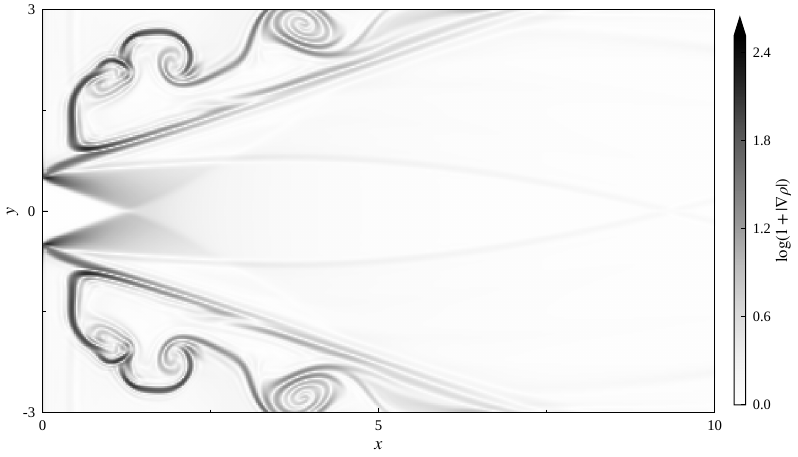}
		\caption{$t=8$.}
		\label{fig:slot-jet-density-gradient-T8}
	\end{subfigure}
	
	\caption{
		Density-gradient indicator for the Mach--3 slot jet at
		$t=2,4,6$, and $8$. The Schlieren-type diagnostic highlights the
		shock-cell pattern, compression fronts, slot-lip shear layers, and
		downstream wave interactions.
	}
	\label{fig:slot-jet-density-gradient-evolution}
\end{figure}

\begin{figure}[h]
	\centering
	\begin{subfigure}[t]{0.48\textwidth}
		\centering
		\includegraphics[width=\linewidth]{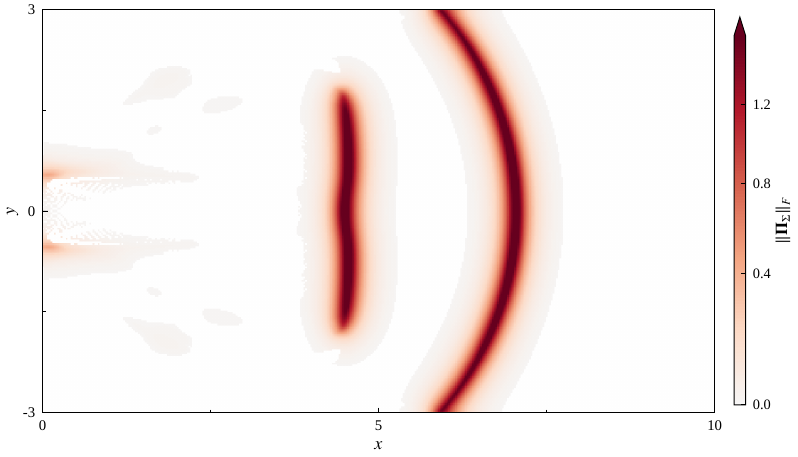}
		\caption{$t=2$.}
		\label{fig:slot-jet-codes-stress-T2}
	\end{subfigure}
	\hfill
	\begin{subfigure}[t]{0.48\textwidth}
		\centering
		\includegraphics[width=\linewidth]{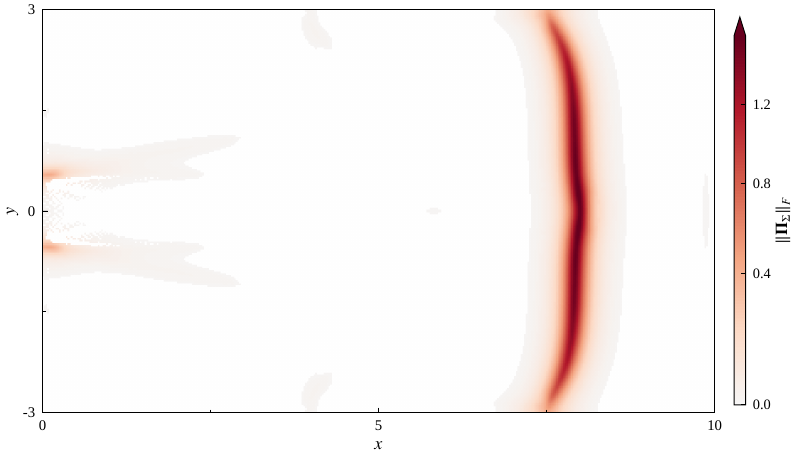}
		\caption{$t=4$.}
		\label{fig:slot-jet-codes-stress-T4}
	\end{subfigure}
	
	\vspace{0.5em}
	
	\begin{subfigure}[t]{0.48\textwidth}
		\centering
		\includegraphics[width=\linewidth]{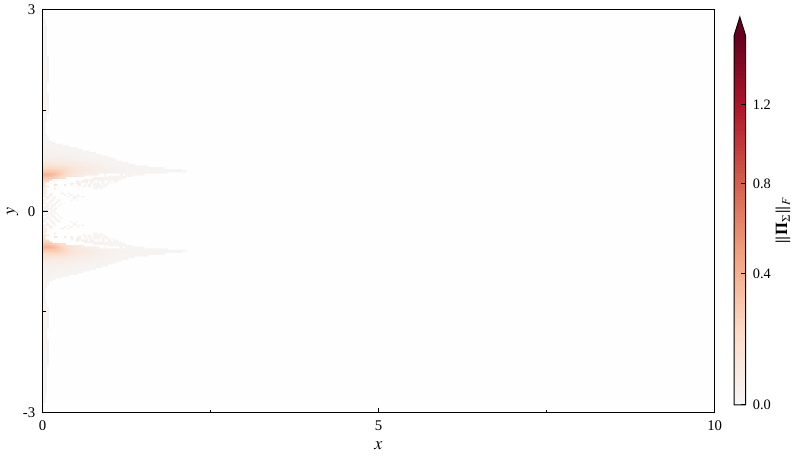}
		\caption{$t=6$.}
		\label{fig:slot-jet-codes-stress-T6}
	\end{subfigure}
	\hfill
	\begin{subfigure}[t]{0.48\textwidth}
		\centering
		\includegraphics[width=\linewidth]{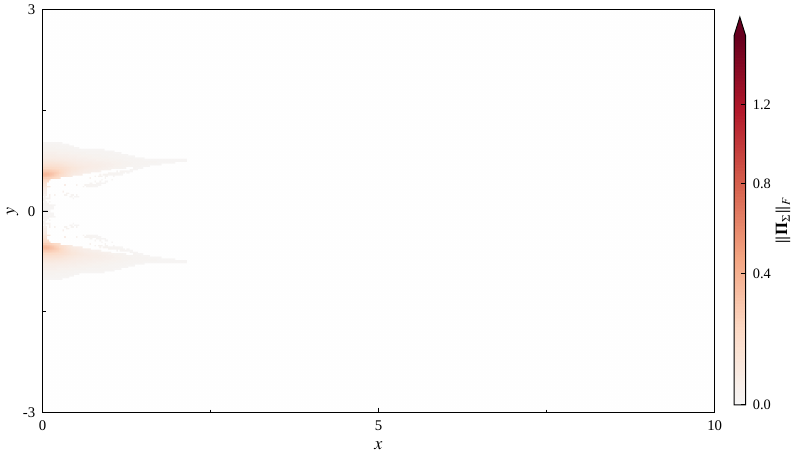}
		\caption{$t=8$.}
		\label{fig:slot-jet-codes-stress-T8}
	\end{subfigure}
	
	\caption{
		CoDeS stress magnitude $\|\boldsymbol{\Pi}_{\Sigma}\|_F$ for the
		Mach--3 slot jet at $t=2,4,6$, and $8$. The stress is concentrated near
		compressive structures, including the inlet compression region, barrel-shock
		system, and downstream shock cells, while remaining weak over much of the
		ambient flow and vortical shear-layer roll-up.
	}
	\label{fig:slot-jet-codes-stress-evolution}
\end{figure}

\begin{figure}[h]
	\centering
	\begin{subfigure}[t]{0.48\textwidth}
		\centering
		\includegraphics[width=\linewidth]{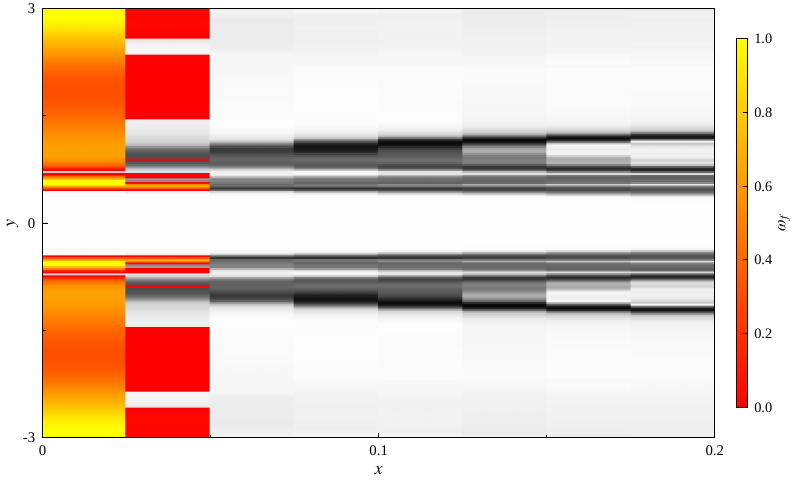}
		\caption{$t=2$.}
		\label{fig:slot-jet-fallback-T2}
	\end{subfigure}
	\hfill
	\begin{subfigure}[t]{0.48\textwidth}
		\centering
		\includegraphics[width=\linewidth]{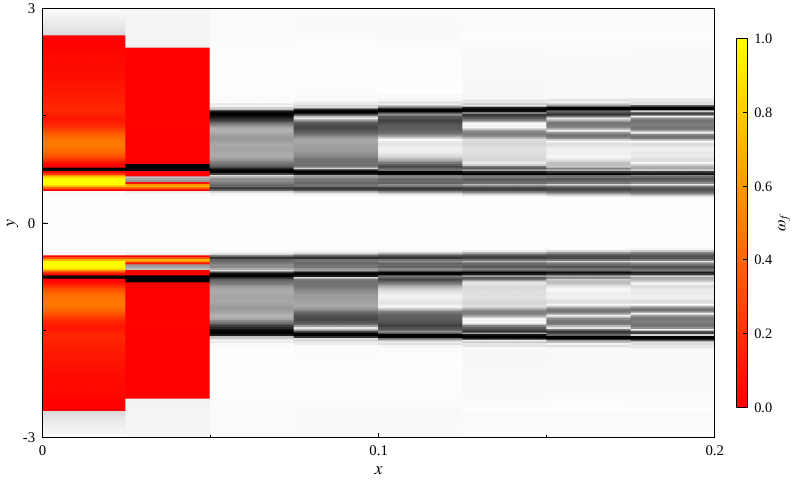}
		\caption{$t=4$.}
		\label{fig:slot-jet-fallback-T4}
	\end{subfigure}
	
	\vspace{0.5em}
	
	\begin{subfigure}[t]{0.48\textwidth}
		\centering
		\includegraphics[width=\linewidth]{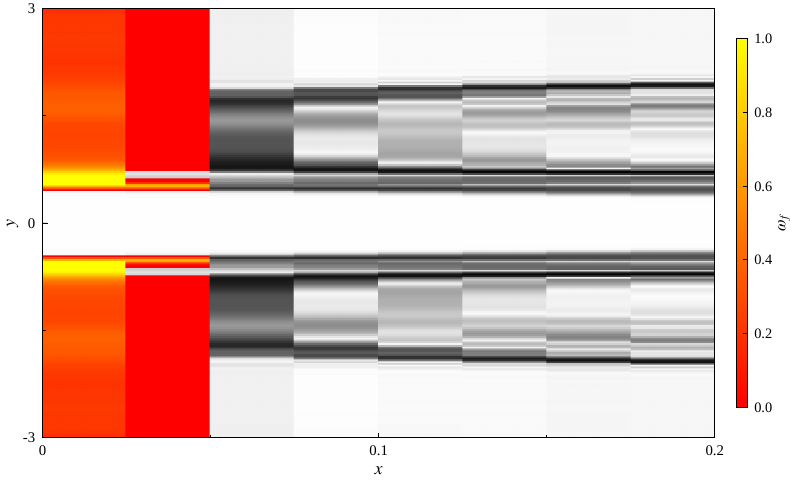}
		\caption{$t=6$.}
		\label{fig:slot-jet-fallback-T6}
	\end{subfigure}
	\hfill
	\begin{subfigure}[t]{0.48\textwidth}
		\centering
		\includegraphics[width=\linewidth]{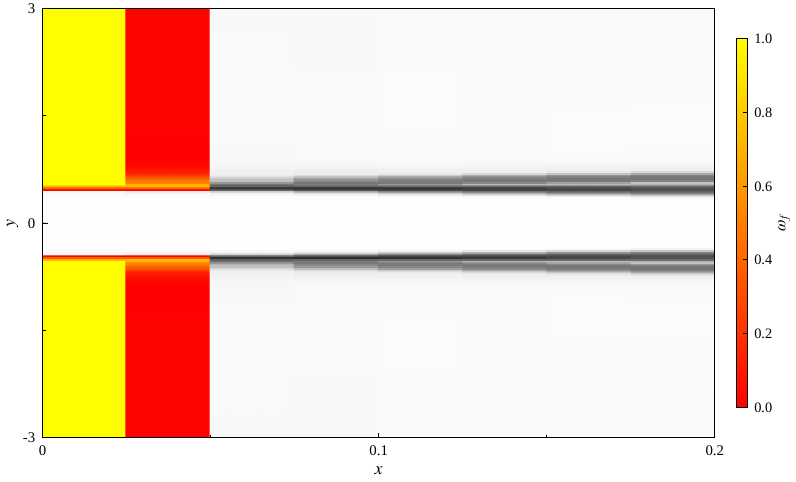}
		\caption{$t=8$.}
		\label{fig:slot-jet-fallback-T8}
	\end{subfigure}
	
	\caption{
		Boundary-localized flux-fallback activation near the left slot-inlet
		boundary at $t=2,4,6$, and $8$. The plotted quantity is the cell-centered
		diagnostic of the maximum adjacent fallback weight $\omega_f$. Activation is
		confined to the prescribed boundary layer near the physical inlet and does
		not replace the high-order flux in the interior jet, shock cells, or
		shear-layer roll-up.
	}
	\label{fig:slot-jet-boundary-fallback}
\end{figure}

The final test is a Mach-3 underexpanded slot jet. This case is intended as a severe robustness test rather than a minimal demonstration of the baseline CoDeS closure. The flow contains a supersonic jet issuing into a quiescent ambient gas, repeated shock-cell formation, barrel-shock and Mach-disk-type compression structures, strong shear layers, and sustained interaction between shocks and nonperiodic physical boundaries. It probes a regime in which interior shock regularization alone may be insufficient, because severe boundary Riemann problems can arise when strong shocks interact with imposed or extrapolated ghost-cell states.

The single-fluid compressible Navier--Stokes equations are solved on $(x,y)\in[0,10]\times[-3,3]$ using a $400\times240$ uniform grid. Viscous terms are retained with $Re=10^6$, so the calculation is effectively a high-Reynolds-number compressible jet dominated by inviscid shock and shear-layer dynamics. Time integration uses SSP-RK3 with adaptive time stepping and target $\mathrm{CFL}=0.12$. The final time is $t=8$.

The ambient state is $(\rho,u,v,p)_a=(1,0,0,1)$, and the jet state is specified by $\rho_j=1$, $p_j=3$, $u_j=M_j\sqrt{\gamma p_j/\rho_j}$, $M_j=3$, and $v_j=0$. The numerical initial condition is smoothed near the slot lip using a tanh function transition. The left boundary at $x=0$ is a Dirichlet slot-inlet boundary, while the right and transverse boundaries use ghost-cell extrapolation.

The CoDeS discretization uses seventh-order reconstruction for both the conservative flux and the stress flux, with $\boldsymbol{\Pi}_{\Sigma}=\sigma\boldsymbol{M}$ and $\alpha=20h^2$. The modified-Helmholtz equation for $\sigma$ is solved using three Jacobi iterations per Runge--Kutta stage after a warm start. Because this case is substantially more severe than the preceding tests, two localized robustness devices are enabled. First, the shock-localized face-normal compression source is activated with $\kappa_f=2.5$. Second, a boundary-localized HLLE flux fallback is applied in one interior face layer adjacent to nonperiodic physical boundaries. The fallback thresholds are $R_{p,0}=1.02$, $R_{p,1}=1.20$, $M_{c,0}=0$, and $M_{c,1}=0.05$, with maximum blend factor equal to one. This fallback acts only on selected Euler fluxes in the prescribed boundary layer and does not modify the CoDeS tensor stress, the modified-Helmholtz solver, or the high-order interior fluxes.

Figure~\ref{fig:slot-jet-density-evolution} shows density fields at $T=2,4,6$, and $8$. At $t=2$, the jet has developed a well-defined core and an initial shock-cell structure. Strong gradient also appears near the inlet-side outer region, where the imposed slot state meets the quiescent ambient gas and generates a severe boundary interaction. By $t=4$, the barrel-shock and downstream compression pattern are fully established. At $t=6$, coherent roll-up structures develop along the outer shear layers while the central jet and shock-cell system remain sharply defined. By $t=8$, the leading compression system has propagated downstream and the plume remains stable, with persistent shear-layer roll-up near the inlet and a coherent downstream density field. The computation remains stable without replacing the interior domain by a globally low-order flux.

The density-gradient indicator in Figure~\ref{fig:slot-jet-density-gradient-evolution} provides a Schlieren-type view of the shock-cell pattern, compression fronts, slot-lip shear layers, and downstream wave interactions. The interior jet continues to be advanced by the high-order finite-volume flux together with the CoDeS stress tensor, and the Schlieren fields confirm that shocks and shear-layer structures remain localized through the time duration.

Figure~\ref{fig:slot-jet-codes-stress-evolution} shows $\|\boldsymbol{\Pi}_{\Sigma}\|_F$. The strongest stress concentrates near compressive structures at the jet inlet, the initial compression fans, the barrel-shock system, and the downstream shock cells. Large portions of the ambient region and much of the vortical shear-layer roll-up produce only weak entropic stress. Thus, even in this severe jet calculation, the regularization is not triggered merely by large gradient or vorticity but remains primarily localized to compressible wave structures. This distinction is essential, because excessive regularization of the shear layers would suppress the roll-up dynamics that form part of the physically relevant jet evolution.

The boundary-fallback activation is shown in Figure~\ref{fig:slot-jet-boundary-fallback} through the stored blend weight $\omega_f$. Activation remains confined to the prescribed physical-boundary layer and is the most visible near the inlet. The plotted diagnostic is cell-centered and stores the maximum fallback weight over adjacent faces, so its visual footprint may span more cells than the actual face set on which the flux fallback is applied. Away from the boundary, the shocks, shear layers, and downstream compression structures are not replaced by a first-order HLLE flux.

The role of the fallback is the clearest at the Dirichlet boundaries near the slot inlet. The imposed supersonic jet and the surrounding ambient state generate large local pressure and normal-velocity jumps at the inlet-side boundary. The fallback blends the high-order Euler boundary flux with HLLE only where these jumps exceed the prescribed smooth thresholds. At $t=2,4,6$, and $8$, the numbers of active cell-centered fallback diagnostics are $714,552,464$, and $556$ respectively, out of $96000$ cells, corresponding to less than $0.75\%$ of the domain. The numbers of cells with $\omega_f>0.5$ are even smaller, namely $128,18,28$, and $206$. The fallback therefore acts as a localized boundary-compatibility device rather than as a global dissipation mechanism.The slot-jet configuration also provides a symmetry check. The setup is mirror-symmetric about $y=0$. At $t=6$, the upper and lower active counts are both $232$, the corresponding weight sums are both $37.039$ to the reported precision, and the relative mirror error in the density field is approximately $10^{-13}$. 

\subsection{Supersonic Taylor-Green Vortex}

We finally consider the three-dimensional supersonic Taylor--Green vortex. This smooth periodic problem rapidly develops small-scale vortical structures, compressive waves, and viscous dissipation. It therefore provides a stringent test of whether a shock-regularized high-order method can preserve the vortical cascade without introducing excessive numerical dissipation.

The compressible Navier--Stokes equations are solved for an ideal gas with $\gamma=1.4$ on the periodic domain $(x,y,z)\in[-\pi,\pi]^3$. The initial condition is the compressible Taylor--Green vortex
\begin{equation}
	u=V_0\sin x\cos y\cos z,\qquad
	v=-V_0\cos x\sin y\cos z,\qquad
	w=0 ,
\end{equation}
with
\begin{equation}
	p=p_0+\frac{\rho_0 V_0^2}{16}
	(\cos 2x+\cos 2y)(\cos 2z+2),
	\qquad
	\rho=\rho_0\frac{p}{p_0}.
\end{equation}
The reference parameters are $\rho_0=1$ and $p_0=101325$. The velocity scale $V_0$ is chosen such that the initial Mach number is $M_0=V_0/\sqrt{\gamma p_0/\rho_0}=1.25$. The Reynolds number is $Re=\rho_0 V_0 L/\mu=1600$ with $L=1$, and time is nondimensionalized by $t_c=L/V_0$.

Three seventh-order calculations are compared. The proposed calculation uses the seventh-order CoDeS finite-volume discretization with the tensor entropic-stress closure and regularization parameter $\alpha=0.25h^2$. The reference calculations are performed with MFC using seventh-order WENO and seventh-order TENO reconstruction, both with the LF/Rusanov flux. The MFC TENO calculation uses $C_T=10^{-6}$. All calculations use the same initial condition, periodic boundary conditions, final time $t/t_c=20$, and output interval $\Delta t/t_c\simeq0.2$. CoDeS is run at $128^3$, $256^3$, and $512^3$ resolution, while the MFC WENO-7/LF and TENO-7/LF comparisons are run at $128^3$ and $256^3$. An MFC TENO-7/LF calculation at $512^3$ was also attempted on the same hardware but exceeded the memory available on four V100-32GB GPUs before time stepping and is therefore not reported.

The primary diagnostic is the nondimensional solenoidal dissipation,
\begin{equation}
	\epsilon_s
	=
	\frac{1}{Re}
	\left\langle
	|\boldsymbol{\omega}^{*}|^2
	\right\rangle,
	\qquad
	\boldsymbol{\omega}^{*}
	=
	\frac{L}{V_0}\nabla\times\boldsymbol{u},
\end{equation}
where $\langle\cdot\rangle$ denotes a volume average. This quantity measures the vortical contribution to the dissipation and is commonly used to assess the turbulent cascade in Taylor-Green vortex calculations.

\begin{figure}[t]
	\centering
	\includegraphics[width=0.7\textwidth]{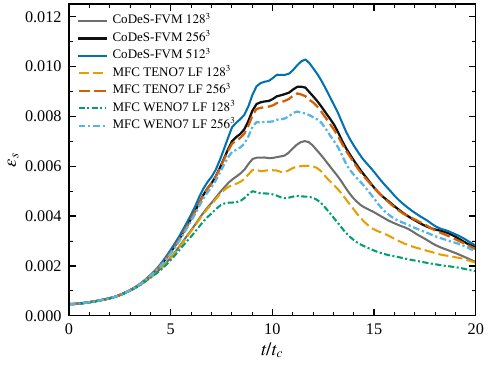}
	\caption{
		Nondimensional solenoidal dissipation $\epsilon_s$ for the supersonic
		Taylor--Green vortex. CoDeS, WENO-7/LF, and TENO-7/LF are compared at the
		available grid resolutions. The CoDeS sequence shows increasing peak
		solenoidal dissipation under refinement, indicating improved resolution of
		the vortical cascade.
	}
	\label{fig:tgv-solenoidal-dissipation}
\end{figure}

Figure~\ref{fig:tgv-solenoidal-dissipation} compares the solenoidal-dissipation histories for all the completed calculations, and Table~\ref{tab:tgv-summary} summarizes the corresponding peak dissipation values, peak times, and end-to-end wall-clock costs. At $128^3$, CoDeS gives a larger solenoidal dissipation peak than both MFC reference schemes, with $7.0024\times10^{-3}$ for CoDeS, compared with $6.0161\times10^{-3}$ for TENO-7/LF and $4.9942\times10^{-3}$ for WENO-7/LF. The WENO-7/LF solution also reaches its peak much earlier, at $t/t_c\simeq9.03$, indicating stronger under-resolved dissipation at this coarse resolution.

At $256^3$, the three methods become closer, but the same ordering remains. CoDeS reaches a peak value of $9.1812\times10^{-3}$, TENO-7/LF reaches $8.9128\times10^{-3}$, and WENO-7/LF reaches $8.1885\times10^{-3}$. Thus, for this case, TENO-7/LF is less dissipative than WENO-7/LF, while CoDeS retains slightly more solenoidal activity than both reference schemes at the same resolution.

The CoDeS resolution sequence further indicates systematic recovery of the vortical cascade. Increasing the resolution from $128^3$ to $256^3$ raises the peak solenoidal dissipation from $7.0024\times10^{-3}$ to $9.1812\times10^{-3}$, and the $512^3$ calculation further increases the peak to $1.0276\times10^{-2}$. The peak time remains near $t/t_c\simeq11.4\text{--}11.6$ for all the CoDeS resolutions. This behavior is consistent with a method that resolves progressively more small-scale vortical activity under mesh refinement, rather than suppressing the cascade through excessive numerical diffusion.

The wall-clock data in Table~\ref{tab:tgv-summary} show that the CoDeS calculation does not incur a prohibitive practical overhead in this three-dimensional benchmark. At $256^3$, CoDeS required 2\,h 23\,min on the four-V100 node, compared with 3\,h 29\,min for TENO-7/LF and 3\,h 18\,min for WENO-7/LF. These timings should be interpreted as practical code-to-code measurements rather than implementation-independent algorithmic costs, because CoDeS and MFC use different code paths and data structures. The relevant point for the present study is that the tensor CoDeS regularization remains practical at three-dimensional scale while producing a solenoidal-dissipation history at least as energetic as the seventh-order WENO/TENO references at the same grid resolution.

\begin{table}[h]
	\centering
	\small
	\caption{
		Summary of peak solenoidal dissipation and wall-clock cost for the
		supersonic Taylor--Green vortex. The peak values are extracted from the
		time histories up to $t/t_c=20$. Wall times are end-to-end timings on the
		same four-V100 GPU node.
	}
	\label{tab:tgv-summary}
	\setlength{\tabcolsep}{6pt}
	\begin{tabular}{@{}l c c c c@{}}
		\toprule
		Method & Grid & Peak $\epsilon_s$ & Peak time $t/t_c$ & Wall time \\
		\midrule
		CoDeS-FVM, 7th & $128^3$ & $7.0024\times10^{-3}$ & 11.585 & 13 min 10.61 s \\
		CoDeS-FVM, 7th & $256^3$ & $9.1812\times10^{-3}$ & 11.425 & 2 h 23 min 23 s \\
		CoDeS-FVM, 7th & $512^3$ & $1.0276\times10^{-2}$ & 11.626 & 29 h 54 min 29 s \\
		\addlinespace
		MFC TENO-7/LF & $128^3$ & $6.0161\times10^{-3}$ & 11.781 & 15 min 56.85 s \\
		MFC TENO-7/LF & $256^3$ & $8.9128\times10^{-3}$ & 11.225 & 3 h 28 min 47 s \\
		\addlinespace
		MFC WENO-7/LF & $128^3$ & $4.9942\times10^{-3}$ & 9.032  & 15 min 36.75 s \\
		MFC WENO-7/LF & $256^3$ & $8.1885\times10^{-3}$ & 11.225 & 3 h 17 min 30 s \\
		\bottomrule
	\end{tabular}
\end{table}

Overall, the supersonic Taylor--Green vortex demonstrates that CoDeS remains effective in a genuinely three-dimensional compressible vortical flow. The method does not stabilize the computation by indiscriminately damping the vortical cascade. The solenoidal dissipation increases systematically with grid refinement, and the $256^3$ CoDeS result retains more vortical activities than the corresponding WENO-7/LF and TENO-7/LF calculations. This supports the intended interpretation of CoDeS as a compression-directed regularization mechanism that remains compatible with high-order resolution of small-scale vortical structures.

\section{Conclusion}\label{sec:conclusion}

We have introduced the Compression-Directional Entropic Stress (CoDeS) method, a tensor regularization for compressible Euler and Navier--Stokes calculations. The method is motivated by the observation that scalar multidimensional IGR-type sources can respond to deformation in a way that is not selective to shock-forming compression. CoDeS retains the modified-Helmholtz smoothing mechanism of IGR but replaces the scalar entropic pressure by a stress tensor whose direction is determined by the compressive eigenspace of the symmetric velocity-gradient tensor. The resulting one-channel closure requires only one scalar modified-Helmholtz solve per Runge--Kutta stage and enters the governing equations through conservative momentum and energy stress fluxes.

The construction was designed to satisfy several structural requirements. In one spatial dimension, CoDeS recovers the compressive part of the scalar IGR entropic-pressure mechanism and therefore retains shock regularization for planar shocks. In pure expansion, pure rigid-body rotation, and ideal contacts with constant velocity and pressure, the source and tensor stress vanish. In smooth compressible flows, the support of the stress tensor in the compressive eigenspace gives a formal nonnegative entropy-production mechanism. These properties distinguish CoDeS from scalar multidimensional regularizations that do not separate compression, expansion, shear, and rotation at the tensor level.

The numerical results support the intended selectivity. In the smooth isentropic simple-wave expansion, CoDeS remains inactive and recovers the high-order behavior of the underlying seventh-order finite-volume discretization, while scalar IGR generates stress in the expansion fan and shows substantially larger errors. In the double-rarefaction problem, CoDeS preserves the rarefaction structure without appreciable stress in the expansive regions. In the Sod shock tube case, the method generates localized stress at the right-going shock while remaining inactive in the rarefaction and at the contact, confirming that the compression gate does not suppress the desired shock regularization.

The multidimensional tests further demonstrate that CoDeS localizes regularization to compressive wave structures without acting as a generic vorticity or shear viscosity. In the two-dimensional isentropic vortex, CoDeS maintains lower long-time pressure error than scalar IGR and remains competitive with WENO-5/LF. In the perturbed two-dimensional Riemann problem, the CoDeS stress concentrates along shocks and strong compression fronts while remaining weak in the central shear-layer roll-up. This yields cleaner shock transitions than WENO-5/LF and avoids the roll-up distortion observed with scalar IGR. In the viscous shock-tube problem, CoDeS regularizes the reflected shock system and oblique compression branches while leaving wall-generated vorticity and shear-layer roll-up largely unaffected. In the two-fluid triple-point calculation, the stress remains concentrated near compressive waves and weak over most of the rolled-up material interface, indicating compatibility with multi-material interface deformation and baroclinic vorticity generation.

The Mach-3 slot-jet calculation provides a severe robustness test involving shock cells, barrel-shock structures, shear-layer roll-up, and strong nonperiodic boundary interactions. In this case, two implementation-level safeguards were enabled, namely a shock-localized face-normal compression source and a boundary-localized HLLE flux fallback. These devices are not part of the continuous CoDeS closure. They modify only the scalar source of the stress-amplitude equation and selected boundary-layer Euler fluxes, respectively. The calculation remains stable through the time duration, and the stress diagnostic shows that CoDeS concentrates near the inlet compression region, barrel-shock system, and downstream shock cells, while the boundary fallback is confined to the prescribed boundary layer.

Finally, the supersonic Taylor--Green vortex demonstrates that CoDeS remains effective and practical in a genuinely three-dimensional compressible vortical flow. The CoDeS resolution sequence shows increasing peak solenoidal dissipation under refinement, indicating progressive recovery of small-scale vortical activity rather than suppression of the cascade by excessive numerical diffusion. At matched resolutions, the CoDeS solenoidal-dissipation histories are at least as energetic as the seventh-order WENO/TENO references considered here, and the measured wall-clock costs do not indicate a prohibitive overhead for three-dimensional calculations.

The present study has several limitations. The entropy-production property is a continuous design principle, not a complete discrete entropy-stability proof or a convergence theorem for the unregularized Euler equations. The finite-volume implementation used here is cell-centered and structured-grid based, and the modified-Helmholtz equation is solved with practical iterative updates rather than an optimized elliptic solver. The parameters controlling the smoothing length and optional safeguards were chosen empirically for the present tests. Future work should therefore address discrete entropy stability, systematic parameter selection, adaptive choices of the regularization length scale, extensions to curvilinear and unstructured meshes, optimized multigrid or preconditioned solvers for the stress amplitude, and coupling with entropy-stable or positivity-preserving high-order finite-volume and discontinuous-Galerkin discretizations.

Overall, the results indicate that CoDeS provides a compression-directional alternative to scalar entropic-pressure regularization. By aligning the stress with the local compressive eigenspace, the method supplies shock regularization where compressive stabilization is needed while reducing false activation in expansion, contacts, shear layers, material interfaces, and vortical flow. This combination makes CoDeS a promising regularization strategy for high-order simulation of multidimensional compressible flows with interacting shocks, interfaces, and turbulence.

\printcredits

\bibliographystyle{elsarticle-num-names}


\bibliography{cas-refs}

\end{document}